\documentclass[preprint,10pt]{elsarticle}
\usepackage[table]{xcolor}
\usepackage[a4paper, margin=2cm]{geometry}

\usepackage{amssymb}
\usepackage{amsmath}

\usepackage{enumitem}
\usepackage{subcaption}
\usepackage{svg}  
\usepackage{graphicx}
\usepackage{float}
\usepackage{makecell}
\usepackage{hyperref}

\begin{document}

\begin{frontmatter}



\title{Multi-fidelity graph-based neural networks architectures to learn Navier-Stokes solutions on non-parametrized 2D domains} 

\author[1]{Francesco Songia}
\author[1]{Raoul {Sallé de Chou}}
\author[1,2]{Hugues Talbot}
\author[1]{Irene E. Vignon-Clementel\textsuperscript{*}}

\address[1]{Inria, Research Center Saclay Ile-de-France, France}
\address[2]{CentraleSupelec, Universit\'{e} Paris-Saclay, France}

\cortext[*]{Corresponding author: \texttt{irene.vignon-clementel@inria.fr}}

\begin{abstract}
We propose a graph-based, multi-fidelity learning framework for the prediction of stationary Navier–Stokes solutions in non-parametrized two-dimensional geometries. The method is designed to guide the learning process through successive approximations, starting from reduced-order and full Stokes models, and progressively approaching the Navier–Stokes solution. To effectively capture both local and long-range dependencies in the velocity and pressure fields, we combine graph neural networks with Transformer and Mamba architectures. While Transformers achieve the highest accuracy, we show that Mamba can be successfully adapted to graph-structured data through an unsupervised node-ordering strategy. The Mamba approach significantly reduces computational cost while maintaining performance.\\
Physical knowledge is embedded directly into the architecture through an \emph{encoding - processing - physics informed decoding} pipeline. Derivatives are computed through algebraic operators constructed via the Weighted Least Squares method. The flexibility of these operators allows us not only to make the output obey the governing equations, but also to constrain selected hidden features to satisfy mass conservation. We introduce additional physical biases through an enriched graph convolution with the same differential operators describing the PDEs. Overall, we successfully guide the learning process by physical knowledge and fluid dynamics insights, leading to more regular and accurate predictions.

\end{abstract}




\end{frontmatter}


\section{Introduction}
Solving partial differential equations, such as the Navier--Stokes (NS) equations, with traditional computational fluid dynamics (CFD) solvers provides accurate velocity and pressure fields in complex domains, typically represented by unstructured meshes. The main drawback is that each simulation can take several hours to days. Deep learning methods offer a way to accelerate this process by learning how to represent these physical fields. Once a neural network is trained on multiple geometries with their corresponding physical solutions, it can predict velocity and pressure fields for previously unseen domains, while respecting the governing equations. This provides a significant speed-up compared to classical solvers and allows generalization across multiple geometries. These fast solvers can finally be applied in clinical applications \cite{pegolotti2024learning}, where a real-time prediction is often required.\\

Deep learning models offer representational capabilities and can be employed to learn specific physical fields directly from data. However, such models typically require large training datasets, which can be difficult to access, and they may struggle to produce physically consistent solutions across diverse geometric configurations. To address these limitations, recent approaches aim to embed mathematical knowledge of the governing physical system directly into the learning process. For instance, Physics-Informed Neural Networks (PINNs), introduced by \cite{raissi2019physics}, add PDE residuals as additional terms in the loss function. This method aims at constraining the solution to lie in a space consistent with the governing physical laws, providing both regularization and generalization capabilities, while reducing the need for ground truth data. \\

Furthermore, non-linear complex fields, such as the solutions of the NS equations, can be challenging to learn directly without an initial approximation. Multi-fidelity approaches improve and facilitate the learning process by structuring the model to approximate the final solution through intermediate steps. The network first learns a low-fidelity representation, such as the Stokes solution, and subsequently learns to handle the non-linear convective terms to predict the full Navier--Stokes solution. Low- and high-fidelity networks can be trained together \cite{howard2023multifidelity, velikorodny2025deep, huang2025multi}, with numerous inexpensively simulated or acquired low-fidelity data points while efficiently employing more costly high-fidelity samples. These attempts to build the model based on computational fluid dynamics principles, are a way to guide the learning process with well-assessed mathematical knowledge. Combining theoretical results with the representation power of deep learning methods is at the core of Scientific Machine Learning and is particularly relevant when modeling complex biomedical systems \cite{ahmadi2025physics}.\\

Graph Neural Networks (GNNs) represent a relevant choice for handling a multiplicity of domains with varying numbers of nodes, exploiting the mesh structure with nodes and their connectivity. Information is spread between nodes, through message-passing architectures \cite{pfaff2020learning} with various applications ranging from fluid dynamics \cite{chen2021graph, gao2024finite} to materials science and chemistry \cite{reiser2022graph}. As graph resolution governs how quickly information propagates, multi-scale approaches \cite{fortunato2022multiscale, garnier2025multi, cao2023efficient} have been developed to learn and combine representations obtained at various levels of resolution.\\
Previous studies have leveraged the structural similarities between message-passing schemes and classical numerical methods, such as the Finite Element method \cite{nastorg2024implicit} or the Finite Volume method \cite{li2025learning}, to impose physical constraints during the training of GNNs. In particular, previous work from Sall{\'e} de Chou et al. \cite{de2024finite} focuses on GNNs to predict myocardial perfusion in 3D domains relying on multi-scale and a Finite Volume informed loss. \\

Transformers, introduced in \cite{vaswani2017attention}, have revolutionized natural language processing and learning capabilities in fluid dynamics applications. Positional encoding and attention scores between all nodes enable the model to inexpensively capture long-range interactions within the domain. Graph Transformers \cite{rampavsek2022recipe, chen2022structure} are then introduced to combine these global relations with local dependencies that are implicitly considered by the graph. Several works with fluid dynamics applications \cite{suk2025deep, suk2024lab, janny2023eagle, garnier2025training, wu2024transolver, jiang2025local, garnier2025graph} benefit from this property, as velocity and pressure show recurrent local patterns that are also influenced by what happens in the entire domain (e.g., boundary conditions, obstacles, bifurcations).\\
This capability of modeling all relationships between nodes for standard Transformers comes at a quadratic computational cost with respect to the number of nodes. As a result, applying Transformers to large graphs or to very long sequences becomes computationally prohibitive, even on high-memory GPUs. To address this limitation, models such as Performers \cite{choromanski2020rethinking} and Exphormers \cite{shirzad2023exphormer} modify the attention mechanism by introducing sparsity, thus reducing the complexity toward a linear scaling with respect to the number of tokens, where a token represents a fundamental element of the input sequence, such as a word in natural language processing or a node in the mesh. Within the sequential modeling framework, recent research has renewed interest in recurrent architectures and State Space Models (SSMs) as efficient alternatives to reduce the computational cost of Transformers. Computational cost and GPU memory requirements become increasingly critical as larger and more realistic problems are considered. ParaRNN \cite{danieli2025pararnn} introduces a parallelizable nonlinear RNN, while Gu and Dao propose Mamba \cite{mamba, mamba2}, a selective state space model. SSMs are an efficient (linear) alternative to attention-based modeling architectures, where the context is encoded in a hidden state that can thus represent global dependencies. Mamba extends this concept by introducing a selection mechanism that controls how each token interacts with and updates the hidden state. During the recurrent scanning process, the information is selectively filtered so that only the most relevant tokens update the global state, resulting in a rich and compact representation of the overall context.\\
However, applying these works, designed for sequence modeling, on non-sequential graphs is not obvious. Graph Mamba architectures have been proposed to integrate GNNs with SSMs, enabling the modeling of global relations. To apply these models, the graph must first be converted into a sequence by defining an order of node visits. In \cite{behrouz2024graph}, this ordering is determined by strategies based on subgraphs and random walks, while \cite{wang2402graph} defines heuristics derived from the degree of the node.\\

In this work, we introduce a novel framework for learning the stationary non-linear Navier-Stokes solutions in non-parametrized 2D domains. Our contributions can be summarized as follows.
\begin{enumerate}
    \item We introduce a novel multi-fidelity model based on graph neural networks in fluid dynamics. Through this multi-fidelity approach, the architecture is designed to iteratively learn the final Navier-Stokes fields, step by step, from successive approximations derived from reduced-order and full Stokes solutions.
    \item GNNs are combined with Transformers and Mamba SSMs, to efficiently capture and integrate both local and global relations within the domain. This is relevant in fluid dynamics applications, where velocity and pressure fields present complex interactions between local and non-local patterns. To be able to apply Mamba to graph-structured data, we propose an unsupervised method to define a transversal order for navigating the graph. Originally developed for sequential data, Mamba overcomes the quadratic computational cost of Transformers, making it a more efficient alternative for large graphs. This feature is important to handle large meshes, necessary for real applications.
    \item We incorporate physical knowledge through an \emph{encoding - processing - physics informed decoding} pipeline. In this formulation, we introduce physics within the architecture itself, not only on the final outputs, by encouraging selected hidden features to obey the governing physical laws. These representations lie in the same functional space as the final velocity and pressure fields, providing physically meaningful features that stabilize and generalize the learning process. From these special features, we introduce additional physical biases through an enriched graph convolution with the same differential operators describing the PDEs. 
\end{enumerate}

\section{Methods}
\label{sec:methods}

\subsection{Multi-fidelity}
\begin{figure}[htbp]
    \centering
    \includegraphics[width=0.85\textwidth]{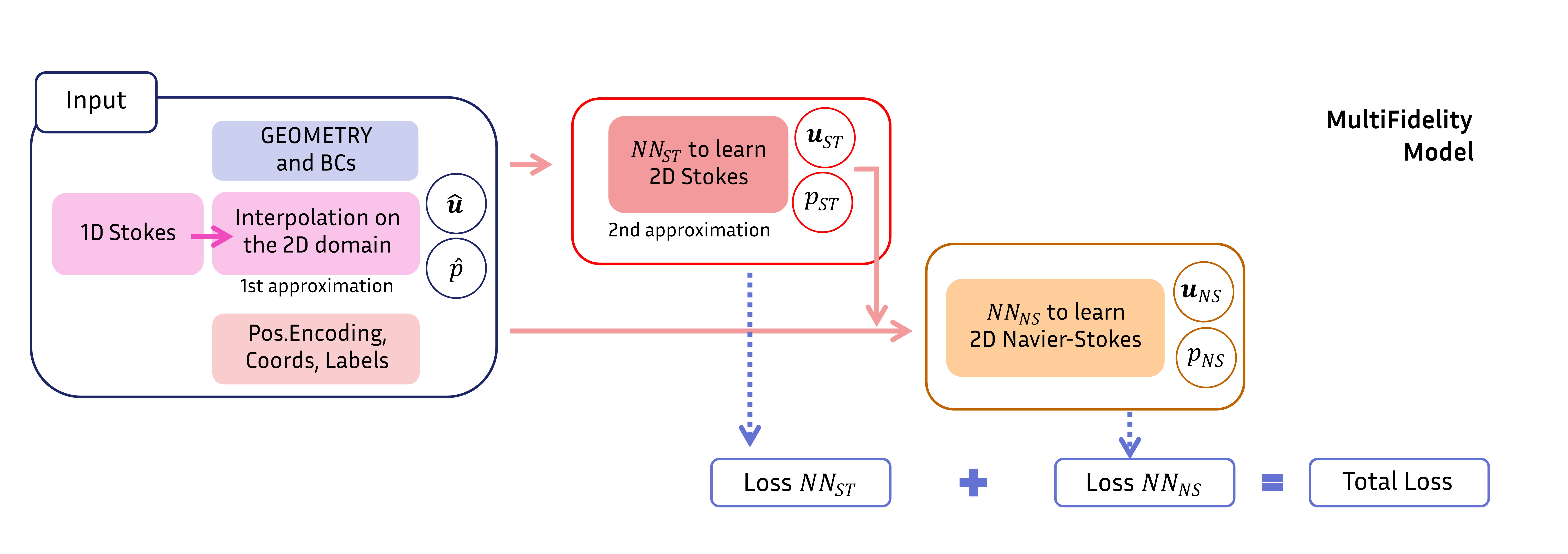}
    \caption{Global multi-fidelity pipeline: two networks are trained together, with the output of the Stokes net that is passed as input for the final Navier-Stokes net.}
    \label{fig:global pipeline}
\end{figure}

The core idea of this work is to progressively learn an approximation of the final NS solution through a combined multi-fidelity and multiscale approach. This strategy begins with 1D centerlines that capture the simpler characteristics of the domain and the flow and gradually advances toward full 2D mesh representations.\\
The final pipeline, represented in Figure \ref{fig:global pipeline}, consists of two neural networks, $\text{NN}_{ST}$ and $\text{NN}_{NS}$, that are trained together following other multi-fidelity approaches \cite{meng2020composite, howard2023multifidelity, huang2025multi}. The first network learns to map the solution of the 1D Stokes equations to the corresponding 2D Stokes solution, denoted as $\mathbf{u}_{ST}$ and $p_{ST}$. The second network then predicts the 2D NS fields, $\mathbf{u}_{NS}$ and $p_{NS}$, from the previously obtained Stokes solution that is concatenated as an additional input.\\
Multiscale is present through the use of reduced-order representations, such as centerlines, together with classical meshes. Multi-fidelity arises from the hierarchy of mathematical models employed to describe the flow: starting from the 1D Stokes equations, moving through the 2D Stokes equations, and finally reaching the 2D NS equations by implicitly learning how to capture the non-linear convective term that characterizes them. The predicted Stokes $(\mathbf{u}_{ST}, p_{ST})$ and Navier--Stokes $(\mathbf{u}_{NS}, p_{NS})$ fields have to satisfy the governing equations reported in (\ref{eq:Stokes2D}) and in (\ref{eq:NS2D}), respectively.
\begin{figure}[H]
\begin{minipage}{0.40\textwidth}
\begin{equation}
\label{eq:Stokes2D}
\left\{
\begin{aligned}
- \mu \, \Delta \mathbf{u}_{ST} + \nabla p_{ST} &= \mathbf{0} &&\text{in } \Omega,\\
\nabla \cdot \mathbf{u}_{ST} &= 0 &&\text{in } \Omega,\\
\mathbf{u}_{ST} &= \mathbf{u}_D &&\text{on } \partial\Omega_{in},\\
p_{ST} &= p_D &&\text{on } \partial\Omega_{out}.
\end{aligned}
\right.
\end{equation}
\end{minipage}
\hfill
\begin{minipage}{0.60\textwidth}
\begin{equation}
\label{eq:NS2D}
\left\{
\begin{aligned}
\rho\,(\mathbf{u}_{NS} \cdot \nabla)\mathbf{u}_{NS}
- \mu \,\Delta \mathbf{u}_{NS}
+ \nabla p_{NS} &= \mathbf{0} &&\text{in } \Omega,\\
\nabla \cdot \mathbf{u}_{NS} &= 0 &&\text{in } \Omega,\\
\mathbf{u}_{NS} &= \mathbf{u}_D &&\text{on } \partial\Omega_{in},\\
p_{NS} &= p_D &&\text{on } \partial\Omega_{out}.
\end{aligned}
\right.
\end{equation}
\end{minipage}
\end{figure}

\subsection{Data}
\label{sec:methods-data}
\begin{figure}[htbp]
    \centering
    \includegraphics[width=\textwidth]{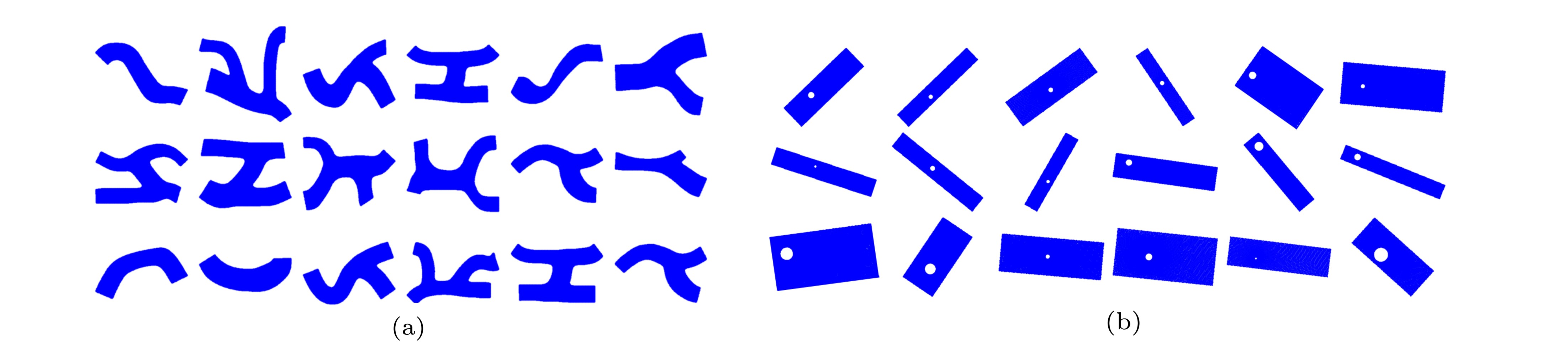}
    \caption{Examples of geometries from (a) {\footnotesize \textit{\textsc{VESSEL}}} and (b) {\footnotesize \textit{\textsc{CYLINDER}}} datasets.}
    \label{fig:shapes}
\end{figure}

We consider two synthetically generated 2D datasets to train and evaluate the models. The former, {\footnotesize \textit{\textsc{VESSEL}}}, mimics real 3D vascular structures with small, short vessels and few bifurcations. The latter, {\footnotesize \textit{\textsc{CYLINDER}}}, represents the classical benchmark of flow around a cylinder. The parameters used to generate the shapes in both cases are never used in the learning process. The number of nodes varies per graph: on average, {\footnotesize \textit{\textsc{VESSEL}}} has $\sim\!7500$ nodes, while {\footnotesize \textit{\textsc{CYLINDER}}} has $\sim\!3500$.  Some examples are shown in Figure \ref{fig:shapes}.\\

For the {\footnotesize \textit{\textsc{VESSEL}}} dataset, simulations were performed using $\mathbb{P}_1$--$\mathbb{P}_1$ elements with SUPG stabilization. The reference physical parameters were set to a density $\rho = 300~\mathrm{kg/m^3}$ and a dynamic viscosity $\mu = 0.005~\mathrm{Pa \cdot s}$. For the {\footnotesize \textit{\textsc{CYLINDER}}} dataset, classical $\mathbb{P}_2$--$\mathbb{P}_1$ elements were employed, with $\rho = 1~\mathrm{kg/m^3}$ and $\mu = 0.001~\mathrm{Pa \cdot s}$.\\
In \ref{app1-data}, we detail the generation of the two datasets, the construction of the initial Stokes--1D approximations, together with the boundary conditions employed.\\

\subsection{Numerical derivatives}
\label{wlsq}
Motivated by the recent success of PINNs, we incorporated physical information in the loss function to enforce the governing PDEs. This requires computing the spatial derivatives of the velocity and pressure fields. In many deep learning applications, automatic differentiation is the standard tool for this purpose. However, with the architectures considered, such as graph neural networks and transformers, the computational graph becomes prohibitively large. For this reason, we adopt an alternative approach inspired by the Weighted Least Squares (WLSQ) method to approximate these derivatives efficiently \cite{zhang2017vertex, white2019weighted, li2025learning}. As we work with graph-based networks, we aim to leverage the structure of nodes and their neighbors to handle derivatives and more complex functionals. This method aligns with classical meshless methods and generalized moving least squares techniques \cite{atluri2002meshless, mirzaei2013direct, le2023guidelines}, with the key difference that our formulation does not rely on predefined basis functions.\\
The core idea is to construct matrices that approximate differential operators, such as the spatial derivatives $\partial(\cdot)/\partial x$, $\partial(\cdot)/\partial y$, or the Laplacian $\Delta (\cdot)$. These matrix operators are precomputed before training and can then be applied directly to any field defined on the graph nodes. With these matrices, we are able to easily compute derivatives of the physical outputs, but also of any latent features.\\

For each point \( i \), we select \( k \) neighboring points and, for each of them, we consider a \( \mathbb{P}^2 \) Taylor expansion around \( (x_i, y_i) \):
\[
u(x_i + \delta x,\, y_i + \delta y) =
u_i
+ \frac{\partial u}{\partial x} \, \delta x
+ \frac{\partial u}{\partial y} \, \delta y
+ \frac{1}{2} \frac{\partial^2 u}{\partial x^2} (\delta x)^2
+ \frac{\partial^2 u}{\partial x \partial y} \, \delta x \delta y
+ \frac{1}{2} \frac{\partial^2 u}{\partial y^2} (\delta y)^2.
\]\\
This expression can be rewritten in compact form by defining the vector of local derivatives \( \boldsymbol{\beta}_i = \)\( [\, u_x,\, \) \(u_y,\, u_{xx}, \ \),\( u_{xy},\,\) \(u_{yy} \,]^{\top} \) and the corresponding local geometric matrix \( \mathbf{A}_i \in \mathbb{R}^{k \times 5} \) that takes into account all the neighbors, such that
\[
u_j - u_i = \mathbf{A}_i \boldsymbol{\beta}_i.
\]
\[
\mathbf{A}_i =
\begin{bmatrix}
\delta x_{j_1} & \delta y_{j_1} & \tfrac{1}{2} \delta x_{j_1}^2 & \delta x_{j_1} \delta y_{j_1} & \tfrac{1}{2} \delta y_{j_1}^2 \\
\delta x_{j_2} & \delta y_{j_2} & \tfrac{1}{2} \delta x_{j_2}^2 & \delta x_{j_2} \delta y_{j_2} & \tfrac{1}{2} \delta y_{j_2}^2 \\
\vdots & \vdots & \vdots & \vdots & \vdots \\
\delta x_{j_k} & \delta y_{j_k} & \tfrac{1}{2} \delta x_{j_k}^2 & \delta x_{j_k} \delta y_{j_k} & \tfrac{1}{2} \delta y_{j_k}^2
\end{bmatrix},
\qquad
\mathbf{u}_j - u_i =
\begin{bmatrix}
u_{j_1} - u_i \\
u_{j_2} - u_i \\
\vdots \\
u_{j_k} - u_i
\end{bmatrix}.
\]\\
To account for the spatial distribution, we consider a distance-based weighting. For each neighbor, we compute \( d_{j} = ((\delta x_j)^2 + (\delta y_j)^2)^{1/2}, \ \text{with} \ \ w_j = \exp\! \ (-(d_j / \sigma)^2), \) and define the diagonal weight matrix \( \mathbf{W}_i = \mathrm{diag}(w_1,\dots,w_k) \). This system can be solved in a least squares approach:
\[
\boldsymbol{\beta}_i = B (u_j - u_i), \qquad
B = \left(\mathbf{A}_i^\top \mathbf{W}_i^2 \mathbf{A}_i\right)^{-1}\mathbf{A}_i^\top \mathbf{W}_i^2 \qquad B \in \mathbb{R}^{5 \times k}.
\]
A selector matrix \( J_i \in \mathbb{R}^{k \times N} \) is introduced to extract node \( i \) and its \( k \) neighboring nodes from the global point cloud, assigning a positive sign to the neighbors and a negative one to the central node. The local derivative vector is then obtained as
\[
\boldsymbol{\beta}_i = B J_i \mathbf{u},
\]
where \( B \) is the local least-squares operator.\\
Finally, we define a local matrix operator \( M_i \in \mathbb{R}^{5 \times N} \) as
\(
M_i = B\, J_i,
\)
so that the vector of local derivatives is obtained as
\(
\boldsymbol{\beta}_i = M_i \mathbf{u}.
\)
The operator \( M_i \) provides the first- and second-order derivatives at point \( \mathbf{x}_i \) when applied to any scalar field \( \mathbf{u} \).  
Starting from \( M_i \), we assemble the global gradient and Laplacian operators \( G_x,\, G_y,\, K \in \mathbb{R}^{N \times N} \) as
\[
\begin{aligned}
G_x(i, :) &= M_i(1, :) \quad &&\text{with } G_x \text{ representing the } \partial / \partial x \text{ operator}, \\[0.3em]
G_y(i, :) &= M_i(2, :) \quad &&\text{with } G_y \text{ representing the } \partial / \partial y \text{ operator}, \\[0.3em]
K(i, :)   &= M_i(3, :) + M_i(5, :) \quad &&\text{with } K \text{ representing the Laplacian } \Delta.
\end{aligned}
\]
In each line, there are \( k+1 \) non-zero entries. In particular, the diagonal entries quantify the contribution of each node 
$i$ to the computation of its own derivative.\\
These three operators rely only on point cloud coordinates, and the price to pay is a small matrix inversion for each point: \(\mathbf{A}_i^\top \mathbf{W}_i^2 \mathbf{A}_i \in \mathbb{R}^{5 \times 5}\), with a total complexity in \(\mathcal{O}(N)\) to invert all \(N\) matrices of a single geometry.\\

With this approach, we directly construct operators to compute the derivatives of any field, avoiding the need to define basis functions and their (simpler) analytical derivatives, as typically done in meshless or finite element methods.\\

\subsection{Architectures}
\label{sec:archi}
At the core of all the network architectures proposed in this work lies the GNNs, which generalizes the convolution operation (originally developed for processing images, e.g., structured grids) to arbitrary graph domains with arbitrary geometries and variable numbers of points.\\
We have discretized the domain with a mesh considered as a graph \( G = (V, E) \). \( V = \{\mathbf{v}_i\}_{i=1}^{N} \) is the set of nodes, and each \( \mathbf{v}_i \) represents the features associated with node \( i \). The set of edges \( E = \{(s_j, r_j)\}_{j=1}^{M} \) defines the connectivity between nodes, where each edge links a sender node \( s_j \) to a receiver node \( r_j \). The learning mechanism in GNNs relies on an iterative message-passing procedure. At layer $l$, each node $i$ is associated with a feature vector $\mathbf{v}_i^{(l)} \in \mathbb{R}^{F_l}$, and the information is exchanged along the edges by aggregating messages from neighboring nodes. A generic message-passing layer is defined as
\[
\mathbf{m}_i^{(l)} = \sum_{j \in \mathcal{N}(i)}
\psi\!\left(\mathbf{v}_i^{(l)}, \mathbf{v}_j^{(l)}, \mathbf{e}_{ij}\right),
\qquad
\mathbf{v}_i^{(l+1)} = \phi\!\left(\mathbf{v}_i^{(l)}, \mathbf{m}_i^{(l)}\right),
\]
where $\mathcal{N}(i)$ denotes the set of neighbors of node $i$, $\mathbf{e}_{ij}$ represents optional edge features, $\psi(\cdot)$ is the message function, and $\phi(\cdot)$ is the update function. Through successive message-passing steps, the information associated with one node progressively propagates across the entire domain. In this way, local interactions are implicitly captured and processed. This locality permits achieving generalization, both across different regions of the same geometry and among distinct geometries, since similar local relations and patterns recurrently appear throughout the dataset.\\

In the following, we first describe the three benchmark models—{\small \textit{\textsc{MeshGraphNet}}}, {\small \textit{\textsc{GNN-UNet}}}, and {\small \textit{\textsc{GraphDeepONet}}}. We then introduce our proposed architectures, {\small \textit{\textsc{GraphTransformer}}} and {\small \textit{\textsc{GraphMamba}}}, which are built upon an \emph{encoding - processing - physics informed decoding} scheme.
The architectures can be used to replace either of the two networks in the global pipeline. In this work, we use the same architecture for both the $\text{NN}_{ST}$ and the $\text{NN}_{NS}$ networks, only the number of parameters changes. All the architectures are designed to have the same inputs and outputs. The input features include the node coordinates, node labels (inlet, outlet, wall, and interior), positional encoding features (Section~\ref{sec:pos-encoding}), and an initial approximation of the solution. The latter corresponds to the 1D Stokes solution for $\text{NN}_{ST}$, and to the Stokes prediction for $\text{NN}_{NS}$.\\
 
\subsubsection{Benchmark graph-based architectures}
\textbf{{\small \textit{\textsc{MeshGraphNet}}}}. We chose as baseline {\small \textit{\textsc{MeshGraphNet}}} proposed by Pfaff et al. in \cite{pfaff2020learning} with its Encode-Process-Decode architecture, which we have re-implemented from scratch. Unlike their original application, our problem does not involve temporal roll-outs. Instead of predicting successive time steps, the network receives an initial approximation of the solution and directly learns to predict the final fields.\\

\textbf{{\small \textit{\textsc{GNN-UNet}}}}. With {\small \textit{\textsc{MeshGraphNet}}}, the information propagates gradually across the domain through successive message-passing steps. However, global relations between distant nodes are not efficiently captured. To start addressing this limitation, we consider another classical architecture, {\small \textit{\textsc{GNN-UNet}}}, which operates across multiple graph resolutions and reduces the distance between remote regions of the domain.\\
First, node features are preprocessed through a graph-convolutional encoder. The resulting encoded representations are then passed through the UNet module, and finally decoded by a graph-convolutional decoder with the same structure as the encoder to produce the velocity components and pressure fields. The UNet is composed of three hierarchical levels: the first corresponds to the full-resolution graph, while subsequent levels are obtained through Self-Attention Pooling \cite{lee2019self, knyazev2019understanding}, which progressively retains half of the nodes at each step. This attention-based pooling mechanism allows the network to learn which nodes are the most relevant to preserve at each resolution level. In each level, graph convolutions are applied to process the node features. Skip connections are included between each level, and a $k$-nearest neighbors (kNN) interpolation is employed to upsample the intermediate representations. \\
With this architecture, the use of multiple resolution levels allows information to propagate more efficiently across the domain, while the self-attention pooling mechanism enables the network to focus on the most relevant nodes. However, global information is still captured only through a sequence of (faster) message-passing steps, and the overall performance remains strongly influenced by the specific choice of the pooling strategy. Furthermore, part of the fine-scale details may be lost during the pooling and unpooling operations.\\

\textbf{{\small \textit{\textsc{GraphDeepONet}}}}. DeepONet was introduced by \cite{lu2019deeponet} as an innovative framework for operator learning, designed to approximate mappings between functions that define a PDE (e.g., external forces, initial and boundary conditions) and its corresponding solution. It has been extended to graph-structured data in \cite{cho2025learning}, where {\small \textit{\textsc{GraphDeepONet}}} learns a mapping from an initial function $u_0$ defined on a set of sensor nodes. Since the cited work also considers time-dependent problems, in our case, we only adopt the architecture idea. We have implemented it from scratch to present an additional comparison on our test cases.  In particular, we define the initial function $u_0$ as the previous approximation in the multi-fidelity framework, together with the node coordinates and additional positional encoding features. This representation is first encoded and then processed by the \emph{branch} network through a series of classical MLP-based message-passing layers. The decoder consists of two separate stages: in the first, the node representations are passed through a soft-attention aggregation mechanism to compute the basis coefficients; in parallel, the node coordinates are processed by the \emph{trunk} network to generate a set of basis functions. The final output is obtained as the dot product between the learned coefficients and the corresponding bases.

\subsubsection{Recover global information}
Small and large scale relations have to be efficiently combined in the final architecture to be able to capture local and global patterns and relations in velocity and pressure fields. The following architectures can capture both these relations thanks to the attention mechanisms or by compressing the context into a state vector.\\ 

\textbf{{\small \textit{\textsc{GraphTransformer}}}}. Transformers are the most powerful choice to learn fields defined on large graphs. They efficiently exchange information between all nodes, effectively capturing local and non-local relations. They can be seen as fully-connected graph neural networks \cite{joshi2025transformers}, and the relevance of each connection can be weighted through a (multi) attention mechanism.\\
Each transformer module is composed of a subsampling, a sequence of TransformerBlocks, and a final kNN interpolation layer. We apply the transformer module on a coarser graph to reduce the computation cost; to choose which nodes to keep, we follow the \emph{PointNet++}  sampling algorithm \cite{qi2017pointnet++}. Inside each TransformerBlock, the latent representation is first normalized with GraphNorm \cite{cai2021graphnorm}, then processed by a multi-head attention (MHA) mechanism \cite{vaswani2017attention}, followed by a second GraphNorm. Finally, the features pass through a GatedMLP \cite{dauphin2017language} with GeLU activation \cite{hendrycks2016gaussian}, as adopted in recent transformer architectures \cite{de2024griffin}. We can summarize the operations of a TransformerBlock as follows:
\[
\begin{aligned}
\mathbf{h_0} &= \text{GraphNorm}_1(\mathbf{h}), \\
\mathbf{h_1} &= \text{GraphNorm}_2\!\left(\mathbf{h_0} + \text{MHA}(\mathbf{h_0})\right), \\
\mathbf{y}   &= \mathbf{h_1}+ \text{GeLU}(W_1 \mathbf{h_1} + b_1) \odot (W_2 \mathbf{h_1}+ b_2).
\end{aligned}
\]

\textbf{{\small \textit{\textsc{GraphMamba}}}}
This architecture is based on Mamba, a structured state space model (SSMs), originally introduced in \cite{mamba} for language processing. The main motivation behind Mamba is to reduce the computational cost of Transformers while still retaining the ability to capture long-range dependencies. Unlike Transformers, which explicitly compute global interactions through attention over all nodes, Mamba gathers global relational structure through its state. It visits the nodes, and it is updated after each step, keeping relevant information through the selection mechanism. By doing so, the state progressively integrates long-range dependencies and thus carries a compact global summary of the graph.\\
In the following, we first recall the mathematical framework of Mamba, then we describe how we adapt it to non-sequential graph data, and finally, we detail the specific Mamba layer employed in our architecture.\\

SSMs are a class of sequence models that map an input sequence 
$\mathbf{x}(t) \in \mathbb{R}^N$ to an output sequence $\mathbf{y}(t) \in \mathbb{R}^N$ via a latent state $\mathbf{h}(t) \in \mathbb{R}^N$. The system can be described in continuous time and discretized time as follows:
\begin{center}
\begin{minipage}{0.45\textwidth}
\[
\left\{
\begin{aligned}
\mathbf{h}'(t) &= A \mathbf{h}(t) + B \mathbf{x}(t), \\
\mathbf{y}(t) &= C \mathbf{h}(t)
\end{aligned}
\right.
\]
\end{minipage}
\begin{minipage}{0.45\textwidth}
\[
\left\{
\begin{aligned}
\mathbf{h}_t &= \bar{A} \mathbf{h}_{t-1} + \bar{B} \mathbf{x}_t, \\
\mathbf{y}_t &= C \mathbf{h}_t
\end{aligned}
\right.
\]
\end{minipage}
\end{center}
Here, $A \in \mathbb{R}^{N \times N}$ and $B, \ C\in \mathbb{R}^{N}$ are the state, input and output state matrices. For the discretized system, $\bar{A} := \exp(\Delta A)$ and $\bar{B} := (\Delta A)^{-1} (\exp(\Delta A) - I) \Delta B$, where $\Delta$ is the discretization step. The initialization of the state matrix $A$ is based on the HIPPO theory \cite{gu2020hippo} to better capture and compress previous tokens visited.\\
State-space models must rely on a finite-dimensional state and are therefore forced to compress contextual information. In the standard setting, the dynamics matrices $\Delta, B, C$ remain fixed over time, limiting the model’s ability to adapt its state to select relevant information. To overcome this, $\Delta, B, C$ become functions of the input \cite{mamba}, enabling input-dependent state transitions that dynamically modulate which information is propagated or suppressed.\\

\begin{figure}[H] 
    \centering
    \includegraphics[width=0.8\textwidth]{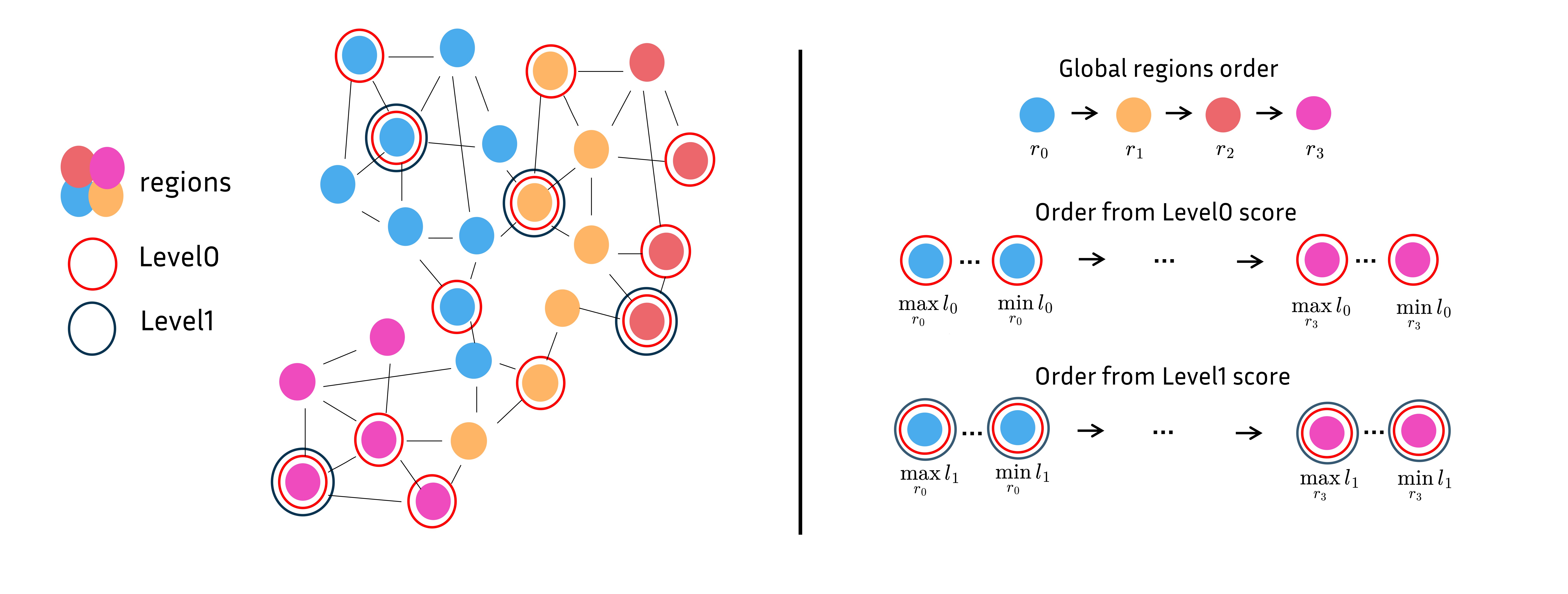}
    \caption{Clustering module with the proposed node ordering based on regions $r$ and levels $l$. There is a first global region order (here, $r_0$, $...$, $r_3$); then we define two orders from the $l_0$ and the $l_1$ scores. Specifically, for the first level, nodes are traversed region-wise, where within each region $r_i$, nodes are visited from the highest to the lowest $l_0$ score. The same logic is applied to the second level $l_1$. }
    \label{fig:clustering}
\end{figure}

Mamba was originally proposed for sequence data that has a natural ordering. A major challenge when applying this model to graphs is to define a node ordering in which the graph is explored. One can define orderings based on node degree \cite{wang2402graph} or through random-walk transversal strategies \cite{behrouz2024graph}. In this work, we propose a Clustering module that learns how to navigate the graph by defining a hierarchy of exploration regions $r$ and levels $l$. With this, we let the network learn an ordering in an unsupervised way. We first compute a global score over all nodes and use it to partition the graph into a fixed number of regions: the nodes with the highest score are assigned to region $r_0$, the next group to region $r_1$, and so on. Importantly, these regions are not required to be connected subgraphs: nodes belonging to the same region may be far apart in the original geometry. \\
To introduce multi-scale exploration, we compute a second score to define the first refinement level $l_0$. For each region, we retain only a ratio $R_0$ of nodes with the highest $l_0$ score. The Mamba update will therefore first visit the $l_0$ nodes of region $r_0$, then the $l_0$ nodes of $r_1$, and so forth. We further refine this ordering by computing an additional score, level $l_1$. Within each region, we retain a ratio $R_1$ of $l_0$ nodes. As for $l_0$, the transversal is again region-wise: the $l_1$ nodes of region $r_0$ are visited first, followed by the $l_1$ nodes of $r_1$, and so on. This construction, represented in Figure \ref{fig:clustering}, defines two global orderings over the graph: one induced by $l_0$ and another induced by $l_1$. We then consider two Mamba modules that traverse the graph along the same sequence of regions but with different transversal speeds. \\
The Clustering module is called at the end of the \emph{encoding} stage, and the same orderings are used for all the processor steps. In the Mamba layer, the latent representation $\mathbf{h}$ is processed by two Mamba blocks that use two computed orderings. First, the coarser graph defined by $l_1$ is processed, the output is then concatenated to the initial $\mathbf{h}$ and goes into the second Mamba block with order induced by $l_0$. Since both orderings are defined on a subgraph, a kNN interpolation is needed between the blocks and to produce the final output defined on all nodes. 

\subsubsection{Introduce physical knowledge within the architecture}
\label{grad-lapl}
\begin{figure}[ht] 
    \centering
    \includegraphics[width=\textwidth]{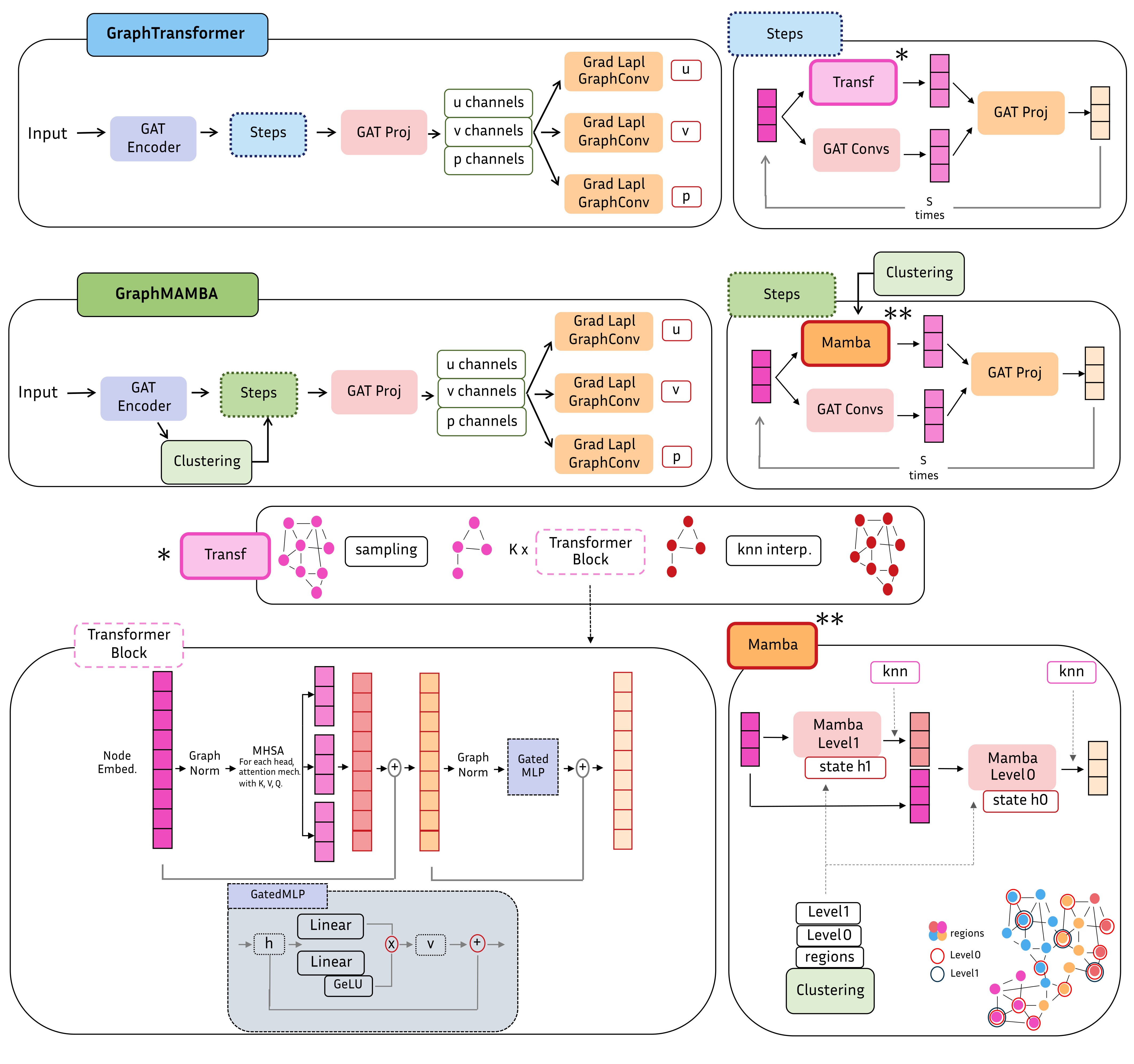}
    \caption{Final {\small \textit{\textsc{GraphTransformer}}} and {\small \textit{\textsc{GraphMamba}}} architectures with the respective steps on the right. Below the Transformer model $(*)$ with its inner Transformer Block, and the Mamba model $(**)$.}
    \label{fig:archis}
\end{figure}

We structure the final {\small \textit{\textsc{GraphTransformer}}} and {\small \textit{\textsc{GraphMamba}}} architectures with an \emph{encoding - processing - physics informed decoding} pipeline. In each of these stages, there are graph-based operations composed of convolutions to process the current latent representation. We refer to these layers collectively as \textsc{GAT}-layers, since they are based on GATv2Conv \cite{brody2021attentive}, where local attention coefficients are added to a classical graph convolution. We describe in the following the proposed pipeline, while in Figure \ref{fig:archis} there are represented the building blocks and the final models.

\begin{itemize}
    \item \emph{Encoding:} a single GAT-based encoder maps the input features 
    $\mathbf{v} \in \mathbb{R}^{N \times d_0}$ into the initial latent representation 
    $\mathbf{h} \in \mathbb{R}^{N \times d}$.
    
    \item \emph{Processing:} the latent representation is iteratively updated through a sequence of processing steps. At each step, the latent representation~$\mathbf{h}$ is updated through two parallel branches: a GAT-based convolution, capturing local interactions, and a {\small \textit{\textsc{GlobalModel}}} responsible for long-range dependencies. We consider {\small \textit{\textsc{GraphTransformer}}} and {\small \textit{\textsc{GraphMamba}}} as possible alternatives for the {\small \textit{\textsc{GlobalModel}}}. Their outputs, both in $\mathbb{R}^{N \times d}$, are concatenated and subsequently projected back to dimension~$d$ through a GAT-based projection layer. This represents the core of the architecture: local and global information are combined, enabling the model to capture both small- and large-scale patterns.

    \item \emph{Physics informed decoding:} the final latent representation $\mathbf{h}$ is mapped into $M$ channel triplets $\{u_{\text{channel}},$ $v_{\text{channel}},$ $p_{\text{channel}}\}$, each defined over all nodes with the same hidden dimension. The final Grad-Lapl Graph Convolution introduces physical biases through its operators and then combines the \emph{channels} to reconstruct the output fields $u, \ v, \ p$. We underline that we are decoding using physical knowledge, since we have built the \emph{channels} to have a physical meaning, as we describe in Section \ref{sec:pde-on-channels}.
\end{itemize}

\begin{figure}[htbp]
    \centering
    \includegraphics[width=0.65\textwidth]{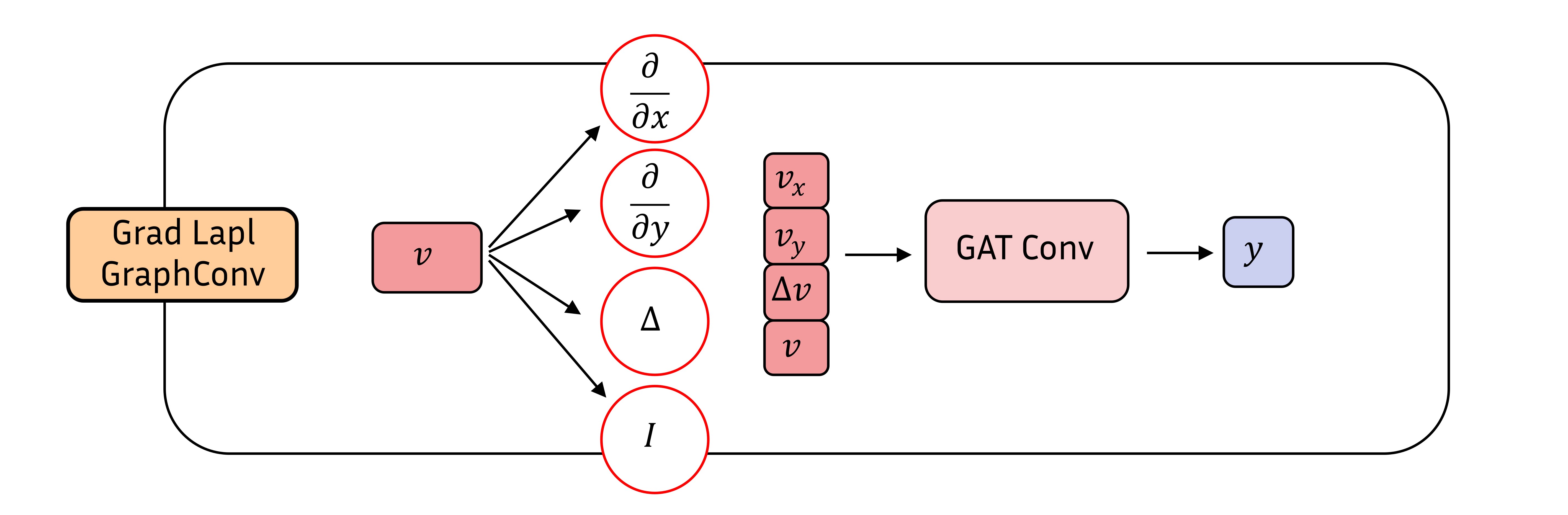}
    \caption{The \emph{Grad-Lapl Graph Convolution} layer. The required derivatives (first order and second order) are computed through the WLSQ operators $G_x, G_y \ \text{and} \ K$. The new features are concatenated and are finally processed by a GAT Convolution layer. }
    \label{fig:grad-lapl}
\end{figure}

\textbf{{Grad-Lapl Graph Convolution}}. To better physically constrain the training, we developed, in the decoder block of {\small \textit{\textsc{GraphTransformer}}} and {\small \textit{\textsc{GraphMamba}}}, the \emph{Grad-Lapl Graph Convolution} module. It is a classical graph convolution where new latent features are added to the input vector. Given an input vector $\mathbf{v}$ defined on the nodes, we compute, using the operators described in Section \ref{wlsq}, the spatial derivatives $\partial \mathbf{v} / \partial x$, $\partial \mathbf{v} / \partial y$, and the Laplacian for each component of $\mathbf{v}$. These new features are normalized through a GraphNorm layer, concatenated with the original input, and finally processed through a standard graph convolution, as illustrated in Figure \ref{fig:grad-lapl}.
This process enriches the latent representation of the node. By doing so, we incorporate physical biases, using operators that are directly relevant to the PDEs governing the system. We define a set of special node features, referred to as \emph{channels}, on which we apply the extended graph convolution. \\
To make these features more informative and closer to the target fields, we can enforce further physical constraints on each \emph{channel} triplet $\{u_\text{channel}, v_\text{channel}, p_\text{channel}\}$. We describe this additional loss term in Section \ref{sec:pde-on-channels}. In this way, the physics-informed \emph{channels} are encouraged to live in a functional space closer to the final outputs, as they obey physical constraints. The gradients and the Laplacians of these quantities are thus expected to be highly informative for the decoding step to predict the final output fields.\\

\subsection{Losses}
For the training of each model, we consider a data-fidelity term and unsupervised physical loss terms. The supervised component, $\mathcal{L}_{\text{sup}}$, is defined as the squared $L^2$-norm with respect to the reference data (\ref{ref-cfd}), and it is always included in all configurations. We introduce additional loss terms to regularize the learning process and enforce the underlying physics of the system. The resulting loss function is defined as:
\[
\mathcal{L} =
\mathcal{L}_{\text{sup}} + \mathcal{L}_{\text{PDE}} + \mathcal{L}_{\text{MASS,channels}}
\]

\subsubsection{Enforce the PDEs on the output}
Starting from the model predictions and using the WLSQ method to compute the derivatives, we compute the residuals over the domain $\Omega$ of the governing equations reported in (\ref{eq:Stokes2D}) and (\ref{eq:NS2D}).\\
First, using the operators derived in Section~\ref{wlsq}, we represent the system of equations defined in $\Omega$, in algebraic form $\mathbf{Ax}=\mathbf{b}$. The right-hand side $\mathbf{b} \in \mathbb{R}^{3N}$ and the system matrix $\mathbf{A} \in \mathbb{R}^{3N \times 3N}$ represent the momentum and mass conservation equations in block form:
\begin{equation*}
\label{eq:ns-matrix}
\begin{bmatrix}
-\mu \ K + \rho \ C(\mathbf{U}) & 0 & G_x \\
0 & -\mu \ K + \rho \ C(\mathbf{U}) & G_y \\
G_x & G_y & 0
\end{bmatrix}
\begin{bmatrix}
U \\ V \\ P
\end{bmatrix}
= 0,
\end{equation*}
\\
where $C(\mathbf{U})$ is the convective term operator evaluated on the current velocity prediction $\mathbf{U}$. For a scalar field $W$, it is defined as $C(\mathbf{U}) \ W = \rho\ U \odot (G_x W) + \rho\ V \odot (G_y W)$. In the following, we illustrate the procedure for computing the NS-based residual loss. The same procedure can be applied to the Stokes equation by removing the convective term.\\
We compute the residuals at each node $i$ and then aggregate them over all $N$ nodes in the graph. If there are multiple domains inside the batch, the residuals are first aggregated on each single graph and then averaged across all graphs. This loss term is finally defined as:
\[
\mathcal{L}_{\text{PDE}} = \alpha \frac{1}{N} \sum_{i=1}^{N} | r_i^{mom_x} | \ \ + \ \  \beta \frac{1}{N} \sum_{i=1}^{N} | r_i^{mom_y} | \ \  + \ \ \gamma \frac{1}{N} \sum_{i=1}^{N} | r_i^{mass} |, 
\]
where each component of the residual is obtained from the block-structured linear system $\mathbf{r} \;=\; \mathbf{b} - \mathbf{A}\mathbf{x}$.\\

\subsubsection{Enforce physical constraints on the \emph{channels}}
\label{sec:pde-on-channels}
In the proposed physics informed decoding stage, we want to introduce physical knowledge not only through the output by minimizing the PDE residuals, but also by regularizing latent features inside the architecture. From each \emph{channel} triplet $\{u_\text{channel}, v_\text{channel}, p_\text{channel}\}$, we take the velocity components, and we enforce mass conservation on each of them by relying on the operators $G_x$ and $G_y$. The goal is to constrain these decoded features to live in the same divergence-free space as the final output.  We define this regularization term as:
\[
\mathcal{L}_{\text{MASS,channels}} = \frac{1}{ N_\text{channels}} \sum_{k=1}^{ N_\text{channels}} \left( \frac{1}{ N}\sum_{i=1}^{ N}  | G_x u_i^{\text{channel}_k} + G_y v_i^{\text{channel}_k}  | \right).
\]
When multiple geometries are present in the batch, the mass–conservation residual is first computed and averaged on each individual graph, then averaged across all graphs, and finally averaged across all \emph{channels}.\\
We enforce only mass conservation to the \emph{channels}, and not the full Navier--Stokes equations, to preserve sufficient freedom in their representation. Enforcing the entire system of equations would instead force the channels to become too similar to each other and to the final output.

\subsection{Training}
Starting from the base geometries, we apply random rotations in the range $[-60^\circ, 60^\circ]$ to increase geometric variability. The corresponding input approximation is rotated accordingly, so that the original $x$- and $y$-directions for the velocity components remain in the reference frame. This procedure also prevents the network from implicitly assuming a privileged flow direction, which is not straightforward to identify in the {\footnotesize \textit{\textsc{VESSEL}}} dataset. It always learns the $x$- and $y$-velocity components with respect to the original axes. With this procedure, we increase the size of the train and test datasets. For {\footnotesize \textit{\textsc{VESSEL}}}, the final dataset consists of 1,700 training samples and 700 test samples, while for {\footnotesize \textit{\textsc{CYLINDER}}}, we obtain 660 training samples and 230 test samples. Each geometry and all its rotated variants are assigned exclusively to the training set or to the test set.\\

We add random Gaussian noise to the input features, excluding the spatial coordinates, to improve robustness and generalization. In all experiments, we use noise distributed as $\mathcal{N}(0,\sigma)$, with $\sigma$ set to $40\%$ of the standard deviation of each feature.\\

We train our models on a single RTX6000 Ada GPU. Particular care is given to GPU memory usage, in order to train our models with sufficient data. For the multi-fidelity model with the two {\small \textit{\textsc{GraphTransformer}}} modules, we enable gradient checkpointing across each Transformer layers at every processing step, which substantially reduces memory consumption during backpropagation. In addition, we employ \texttt{bfloat16} autocasting during training.\\
The architectures are implemented using \texttt{PyTorch Geometric}, the Mamba architecture is implemented in \texttt{PyTorch} using \texttt{mamba-ssm} from \cite{mamba}.

\subsubsection{Positional encoding input features}
\label{sec:pos-encoding}
\begin{figure}[h] 
    \centering
    \includegraphics[width=0.7\textwidth]{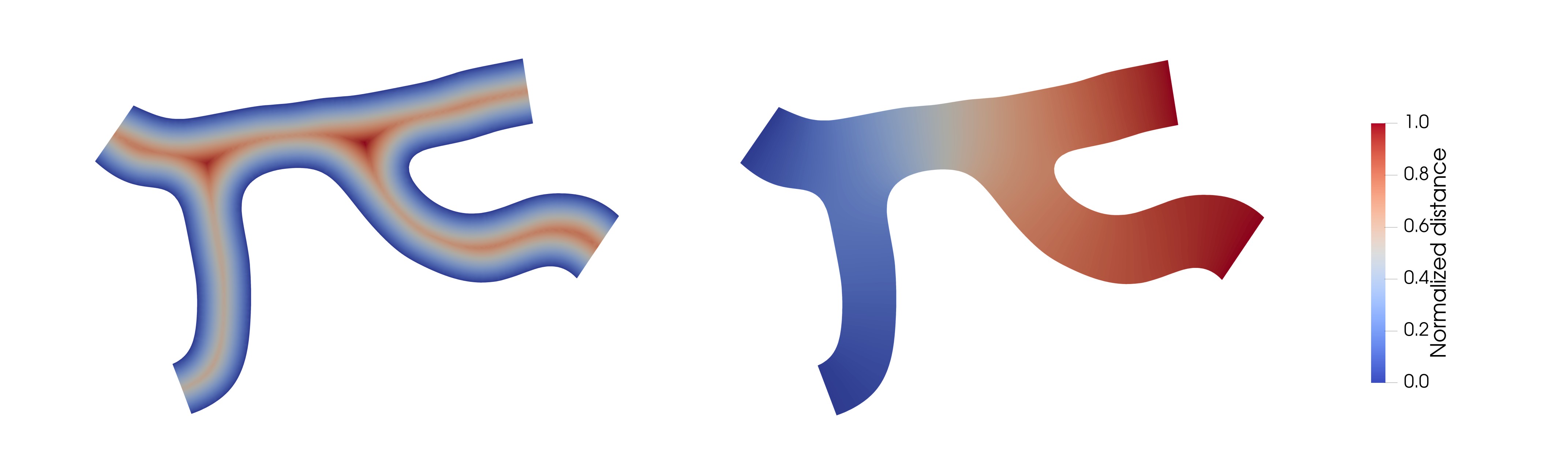}
    \caption{Normalized distance from the wall (left) and \emph{diffusive} distance from the inlets to the outlets (right).}
    \label{fig:pos_encoding}
\end{figure}

As additional input features, we encode the position of each node in the domain. This is similar to what is done in \cite{suk2024deep}, where a geometry-aware location descriptor is employed to encode the position with respect to the inlets and the outlets. Thanks to this positional encoding, Transformers and Mamba can better exchange global information between faraway nodes, as each of them is geometrically characterized in the domain.\\
In particular, we compute the sign distance function from the wall and, to encode the position with respect to the inlets and the outlets, we solve a homogeneous discrete Laplacian problem, imposing a value of $1$ at the outlet boundary and $0$ at the inlet. We represent these two distances in Figure \ref{fig:pos_encoding}. The solution to the discrete Laplacian problem represents a \emph{diffusive} distance that spreads from the inlet to the outlet. We compute this distance in a pre-processing step directly using the mesh structure. In particular, let $\mathbf{A}$ be the connectivity matrix representing the edges; then $\mathbf{A}^\top \mathbf{A}$ defines a discrete Laplacian operator \cite{grady2010discrete}. By imposing the boundary conditions in the corresponding entries of the vector $\mathbf{b}$, we obtain the distance $\mathbf{g}$ by solving the linear system
\[
\mathbf{A}^\top \mathbf{A} \, \mathbf{g} = \mathbf{b}.
\]
This approach is computationally feasible in the 2D case considered here, while for higher-dimensional domains, alternative methods such as the heat method \cite{crane2017heat} could be employed.

\subsubsection{Hyperparameters}
\label{sec:hyperparam}
To train all models, we employed the AdamW optimizer (learning rate $5 \times 10^{-4}$, weight decay $10^{-2}$) together with a cosine annealing scheduler. Training was performed for 800 epochs using a batch size of 16.\\
The multi-fidelity model consists of two networks (Stokes, Navier--Stokes), and we report the main hyperparameters for both.\\
For {\small \textit{\textsc{GraphTransformer}}}, we use latent hidden dimensions $(69, 105)$, $(3, 3)$ processing steps with $(1, 1)$ transformer blocks and $(3, 3)$ attention heads. In the processing step, 40\% of the nodes are sampled to form a coarser graph. For {\small \textit{\textsc{GraphMamba}}}, we use hidden dimensions $(60, 82)$, $(2, 2)$ processing steps. The Mamba state has dimensions $(50, 72)$, and the kernel dimension \emph{dconv} is set to 1. The Clustering module identifies 8 regions, and at each refinement level, half of the nodes are retained ($R_0 = R_1 = 0.5$). For both architectures, we consider $(5, 10)$ \emph{channels} per component, which are concatenated before passing through the \emph{Grad-Lapl Graph Convolution}.\\
For {\small \textit{\textsc{MeshGraphNet}}}, we use hidden dimensions $(55, 70)$ with $(10, 13)$ processing steps. In {\small \textit{\textsc{GNN-UNet}}}, we consider hidden dimensions $(80, 100)$ with $(10, 12)$ graph-convolution layers per level. Finally, for {\small \textit{\textsc{GraphDeepONet}}}, the hidden dimension has been set to $(75, 90)$, there are $(3,3)$ MLP-layers, $(6,8)$ message passing steps, and we consider $(15, 30)$ bases for each output field. \\
The number of learnable parameters for each model is reported in Table~\ref{tab:params}.
\begin{table}[h!]
\centering
\begin{tabular}{lccccc}
\hline
 & {\footnotesize \textit{\textsc{MeshGraphNet}}} & {\footnotesize \textit{\textsc{GNN-UNet}}} & {\footnotesize \textit{\textsc{GraphDeepONet}}} & {\footnotesize \textit{\textsc{GraphTransformer}}} & {\footnotesize \textit{\textsc{GraphMamba}}} \\
\hline
Params Stokes ($k$) & 226 & 254 & 245 & 179 &  231 \\
Params NS ($k$)    &  469 & 454 & 460 & 413  &  438 \\
Total Params ($k$) &  695 &  708 & 705 & 592 & 669 \\
\hline
\end{tabular}
\caption{Number of learnable parameters for each model in the multi-fidelity pipeline, $k$ refers to thousands.}
\label{tab:params}
\end{table}

The weights of the loss terms are manually tuned so that the three supervised components ($u, v, p$) and the three PDEs residuals (mass, $x$- and $y$-momentum) contribute equally to the total loss. The additional term enforcing mass conservation on the \emph{channels} is set to be about one-sixth of the other terms.

\section{Results}
\label{sec:results}
In this Section, we evaluate the proposed multi-fidelity strategy for learning the solution of the NS equations. We compare the five architectures reported in Section \ref{sec:archi} with different loss configurations.
In particular, we evaluate how we can improve the learning process by introducing physical knowledge through an \emph{encoding - processing - physics informed decoding} pipeline. The physics can contribute at three different levels: by enforcing PDE residuals on the final output, by constraining special \emph{channels} to lie in a divergence-free space, and by further incorporating physical biases through a \emph{Grad-Lapl Graph Convolution} acting on these special features. Finally, we evaluate the Mamba SSM from an accuracy point of view, and we compare the computational costs with respect to the Transformer-based architecture.\\

We evaluate the proposed models and configurations on the two datasets defined in Section \ref{sec:methods-data}. Separate trainings are performed on the two datasets. We have performed more tests and comparisons on the {\footnotesize \textit{\textsc{VESSEL}}} dataset as it is the most challenging one, and closer to the real application we are aiming at. The {\footnotesize \textit{\textsc{CYLINDER}}} dataset represents a more classical benchmark.\\
We evaluate predictions with the standardized mean absolute error (SMAE) for the velocity magnitude (VM-SMAE) and pressure (P-SMAE). The Total-SMAE is the sum of the two SMAEs. We normalize by the standard deviation, as it provides a measure of the variability within the geometries. This metric is more informative for datasets characterized by heterogeneous or highly variable regions, such as the bifurcation areas for pressure.\\
For each graph $g$, let $y^{(g)}$ denote the ground truth values and $\hat{y}^{(g)}$ the corresponding predictions. The standardized mean absolute error (SMAE) is defined as the graph-wise mean absolute error, normalized by  
$\bar{\sigma}$, the average standard deviation computed across all graphs:
\[
\mathrm{SMAE} =
\frac{1}{G}\sum_{g=1}^{G} 
\frac{\frac{1}{N_g}\sum_{i=1}^{N_g} \lvert y^{(g)}_i - \hat{y}^{(g)}_i \rvert}
{\bar{\sigma}},
\qquad
\bar{\sigma} = \frac{1}{G}\sum_{g=1}^{G} \mathrm{std}(y^{(g)}) .
\]
These metrics are computed on the test set, which contains 700 samples for {\footnotesize \textit{\textsc{VESSEL}}} and 230 samples for {\footnotesize \textit{\textsc{CYLINDER}}}.\\

\begin{figure}[ht] 
    \centering
    \includegraphics[width=0.95\textwidth]{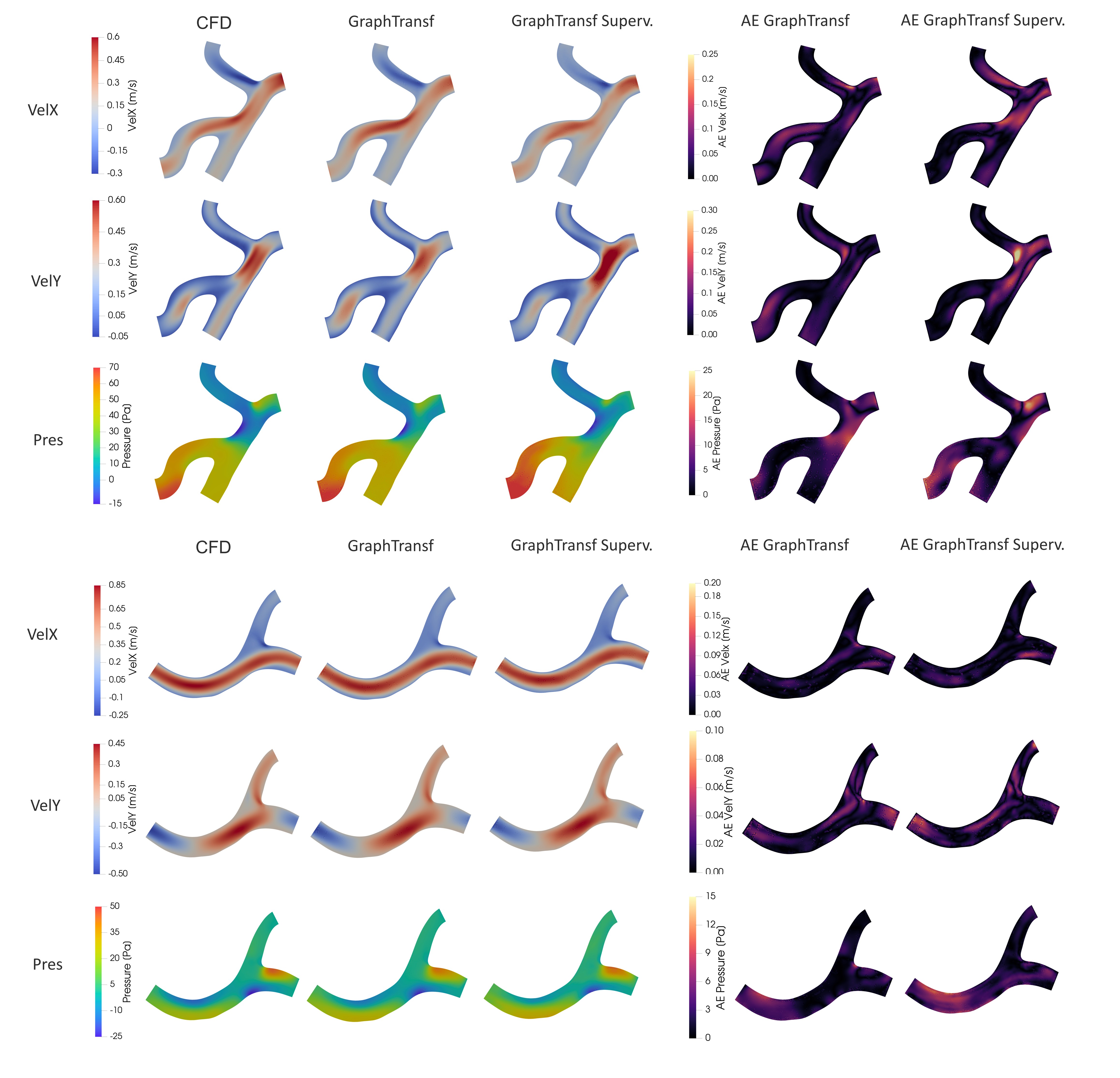}
    \caption{Left: CFD ground truth (1rst column) and predictions of {\small \textit{\textsc{GraphTransformer}}} on the{\footnotesize \textit{\textsc{VESSEL}}} test dataset. 2nd column: the best configuration with mass conservation enforced on the \emph{channels}. 3rd column: comparison with the purely supervised configuration. Right: absolute error maps. The top row geometry belongs to the worst 10\% of the dataset in terms of accuracy.}
    \label{fig:trs_tr}
\end{figure}

\begin{figure}[ht] 
    \centering
    \includegraphics[width=0.95\textwidth]{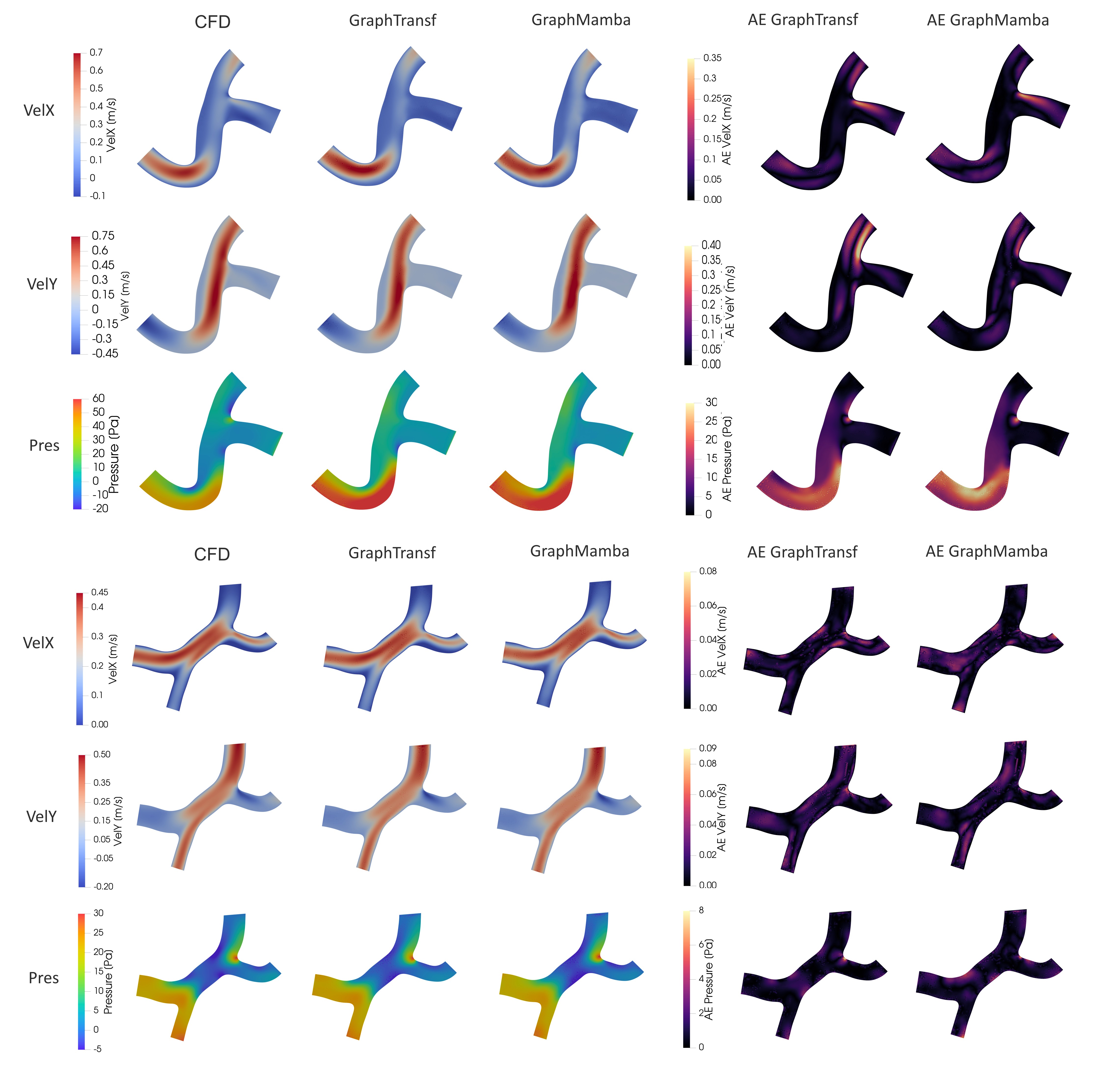}
    \caption{Left: CFD ground truth (1rst column) and predictions of {\small \textit{\textsc{GraphTransformer}}} (2nd column) and of {\small \textit{\textsc{GraphMamba}}} (3rd column) on the {\footnotesize \textit{\textsc{VESSEL}}} test dataset. They both refer to the best loss configuration with mass conservation enforced on the \emph{channels}. Right: absolute error maps. The top row geometry belongs to the worst 10\% of the dataset in terms of accuracy.}
    \label{fig:ma_tr_pdemass}
\end{figure}

\begin{figure}[ht] 
    \centering
    \includegraphics[width=0.95\textwidth]{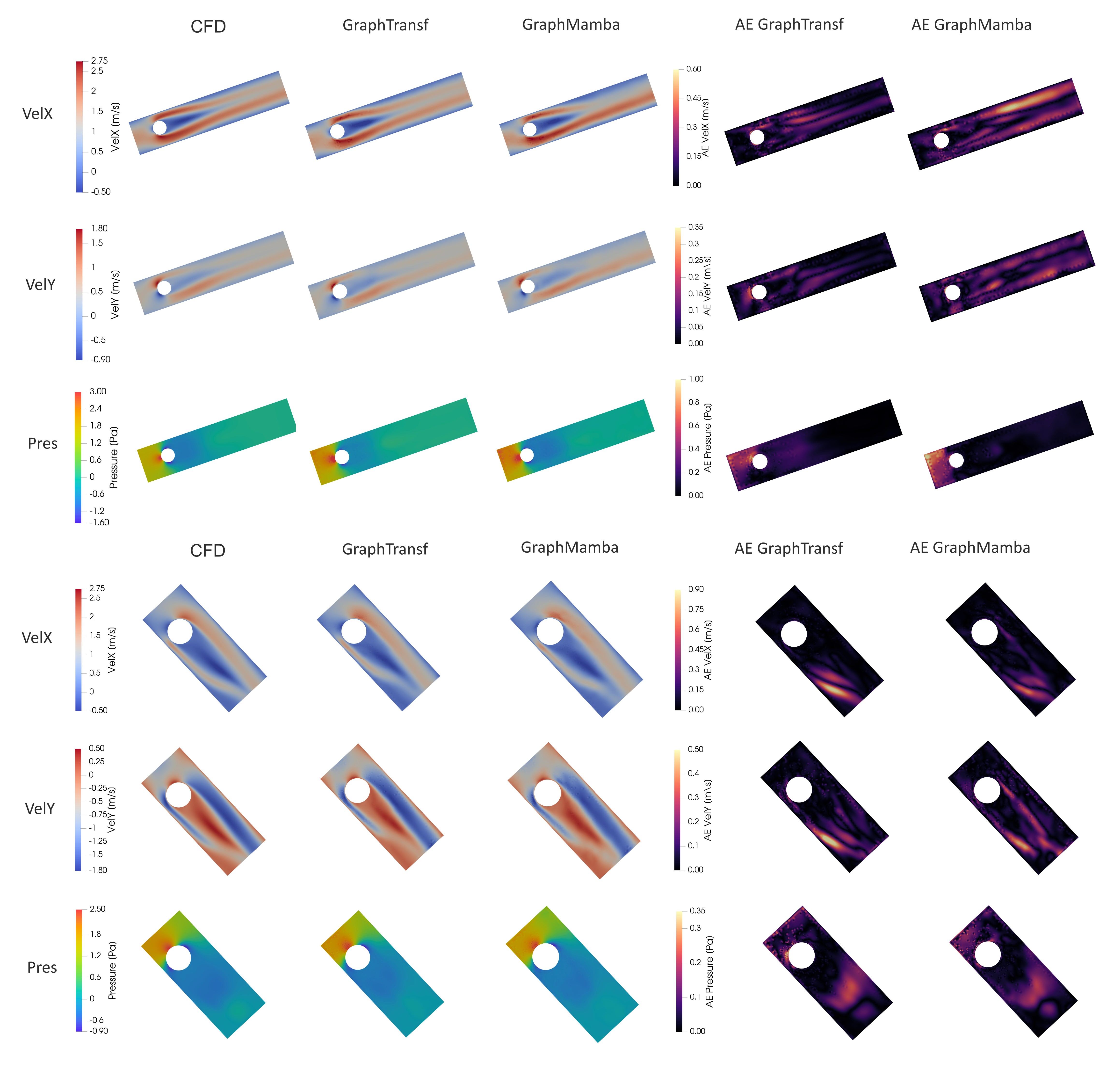}
    \caption{Left: CFD ground truth (1rst column) and predictions of {\small \textit{\textsc{GraphTransformer}}} (2nd column) and of {\small \textit{\textsc{GraphMamba}}} (3rd column) on the {\footnotesize \textit{\textsc{CYLINDER}}} test dataset. They both refer to the best loss configuration with mass conservation enforced on the \emph{channels}. On the right, the absolute error maps.}
    \label{fig:ma_tr_pdemass_cyl}
\end{figure}

We present qualitative visualizations and error maps for selected geometries from the {\footnotesize \textit{\textsc{VESSEL}}} and {\footnotesize \textit{\textsc{CYLINDER}}} datasets, where we compare different model configurations. About {\footnotesize \textit{\textsc{VESSEL}}}, in Figure~\ref{fig:trs_tr}, we compare the purely supervised version of {\small \textit{\textsc{GraphTransformer}}} with the one where mass conservation is enforced on the \emph{channels}; while in Figure~\ref{fig:ma_tr_pdemass}, there are comparisons between {\small \textit{\textsc{GraphMamba}}} and {\small \textit{\textsc{GraphTransformer}}}. We report the same comparison between the Mamba and Transformer network in Figure~\ref{fig:ma_tr_pdemass_cyl} for the {\footnotesize \textit{\textsc{CYLINDER}}} dataset.\\
Additional visualizations are reported in \ref{app2-addit-visualization}, in Figure \ref{fig-app2-ves} and Figure \ref{fig-app2-cyl}.\\

\subsection{Learning local and global relations: model comparison}
{\small \textit{\textsc{GraphTransformer}}} and {\small \textit{\textsc{GraphMamba}}} show better performance compared to the other models, as reported in Tables \ref{tab:res-vessel} and \ref{tab:res-cyl}. In fluid dynamics applications, velocity and pressure fields exhibit both local structures and long-range interactions. Global attention mechanisms, as in Transformers, or compact state representations that encode the overall context, as in Mamba, provide effective ways to model these long-range dependencies. Among the tested architectures, {\small \textit{\textsc{GraphTransformer}}} achieves the best overall results.\\

\subsection{Physics informed loss terms}
Including mathematical knowledge of the physics in the loss function leads to improved results, as shown in Table \ref{tab:res-vessel} and in Table \ref{tab:res-cyl}. While enforcing the governing PDEs only on the final output does not produce a significant difference, imposing mass conservation on special hidden features provides the largest performance gain. By defining physics-informed \emph{channels} that lie in the same divergence-free functional space as the output fields, the model is informed with more meaningful features.\\
\begin{table}[h]
\centering
\begin{tabular}{lccc|c}
\multicolumn{5}{l}{{\footnotesize \textit{\textsc{VESSEL}}} dataset} \\[1pt] 
\Xhline{2pt}
{\footnotesize \textbf{Model}} & {\footnotesize \textbf{Loss}} & {\footnotesize \textbf{VM-SMAE}} & {\small\textbf{P-SMAE} }& {\footnotesize \textbf{Total-SMAE} }\\
\hline
\hline
{\small \textit{\textsc{MeshGraphNet}}} & {\small $\mathcal{L}_{\text{sup}}$ } & 0.404 & 0.662 & 1.066 \\
{\small \textit{\textsc{GNN-UNet}}}   & {\small$\mathcal{L}_{\text{sup}}$ } & 0.407 & 0.662 & 1.069 \\
{\small \textit{\textsc{GraphDeepONet}}}           & {\small $\mathcal{L}_{\text{sup}}$ } & 0.387 & 0.677 & 1.064 \\
\rowcolor{cyan!15}
{\small \textit{\textsc{GraphTransformer}}} &{\small $\mathcal{L}_{\text{sup}}$ }&  0.206 & 0.331 & 0.537  \\
\rowcolor{green!15}
{\small \textit{\textsc{GraphMamba}}}     & {\small$\mathcal{L}_{\text{sup}}$ }& 0.229 & 0.358 & 0.587 \\
\hline
\hline
{\small \textit{\textsc{MeshGraphNet}}}            & {\small $\mathcal{L}_{\text{sup}} + \mathcal{L}_{\text{PDE}}$ } &0.3605 & 0.613 & 0.973  \\
{\small \textit{\textsc{GNN-UNet}}}           & {\small $\mathcal{L}_{\text{sup}} + \mathcal{L}_{\text{PDE}}$ } & 0.406 & 0.633 & 1.039 \\
{\small \textit{\textsc{GraphDeepONet}}}           & {\small $\mathcal{L}_{\text{sup}} + \mathcal{L}_{\text{PDE}}$ } & 0.384 & 0.622 & 1.006 \\
\rowcolor{cyan!15}
{\small \textit{\textsc{GraphTransformer}}} & {\small $\mathcal{L}_{\text{sup}} + \mathcal{L}_{\text{PDE}}$ } & 0.195 & 0.342 & 0.537 \\
\rowcolor{green!15}
{\small \textit{\textsc{GraphMamba}}}     & {\small $\mathcal{L}_{\text{sup}} + \mathcal{L}_{\text{PDE}}$} & 0.227 & 0.354 & 0.581 \\
\hline
\multicolumn{5}{l}{{\small \underline{\emph{With mass conservation on channels}}}} \\[1pt] 
\hline
\rowcolor{cyan!15}
{\small \textit{\textsc{GraphTransformer}}} & {\small$\mathcal{L}_{\text{sup}} + \mathcal{L}_{\text{PDE}} + \mathcal{L}_{\text{MASS,channels}}$ }& 0.195 & 0.317  &  \textbf{0.512}\\
\rowcolor{green!15}
{\small \textit{\textsc{GraphMamba}}}     & {\small $\mathcal{L}_{\text{sup}} + \mathcal{L}_{\text{PDE}} + \mathcal{L}_{\text{MASS,channels}}$ }& 0.223 & 0.355 & 0.578  \\
\hline
\end{tabular}
\caption{Models and losses performance comparison on the {\footnotesize \textit{\textsc{VESSEL}} dataset. VM stands for velocity magnitude and P for pressure contributions.}}
\label{tab:res-vessel}
\end{table}

\begin{table}[h]
\centering
\begin{tabular}{lccc|c}
\multicolumn{5}{l}{{\footnotesize \textit{\textsc{CYLINDER}}} dataset} \\[1pt] 
\Xhline{2pt}
{\footnotesize \textbf{Model}} & {\footnotesize \textbf{Loss}}  & {\footnotesize \textbf{VM-SMAE}} & {\small\textbf{P-SMAE} }& {\footnotesize \textbf{Total-SMAE} }\\
\hline
\hline
{\small \textit{\textsc{MeshGraphNet}}} & {\small $\mathcal{L}_{\text{sup}}$ } & 0.488& 0.480&  0.968\\
{\small \textit{\textsc{GNN-UNet}}}   & {\small$\mathcal{L}_{\text{sup}}$ }& 0.501& 0.453& 0.954\\
{\small \textit{\textsc{GraphDeepONet}}}           & {\small $\mathcal{L}_{\text{sup}}$ } & 0.385 & 0.369 & 0.754\\
\rowcolor{cyan!15}
{\small \textit{\textsc{GraphTransformer}}} &{\small $\mathcal{L}_{\text{sup}}$ }& 0.165& 0.168& 0.333\\
\rowcolor{green!15}
{\small \textit{\textsc{GraphMamba}}}     & {\small$\mathcal{L}_{\text{sup}}$ }& 0.133& 0.140& 0.273\\
\hline
\multicolumn{5}{l}{{\small \underline{\emph{With mass conservation on channels}}}} \\[1pt] 
\hline
\rowcolor{cyan!15}
{\small \textit{\textsc{GraphTransformer}}} & {\small$\mathcal{L}_{\text{sup}} + \mathcal{L}_{\text{PDE}} + \mathcal{L}_{\text{MASS,channels}}$ }& 0.129& 0.131& \textbf{0.260}\\
\rowcolor{green!15}
{\small \textit{\textsc{GraphMamba}}}     & {\small $\mathcal{L}_{\text{sup}} + \mathcal{L}_{\text{PDE}} + \mathcal{L}_{\text{MASS,channels}}$ }& 0.133& 0.136& 0.269\\
\hline
\end{tabular}
\caption{Models and losses performance comparison on the {\footnotesize \textit{\textsc{CYLINDER}} dataset. VM stands for velocity magnitude and P for pressure contributions.}}
\label{tab:res-cyl}
\end{table}

Finally, physical knowledge can be introduced through the \emph{Grad–Lapl Graph Convolution}. To evaluate its effect, we have trained {\small \textit{\textsc{GraphTransformer}}} and {\small \textit{\textsc{GraphMamba}}} without it. The physical biases introduced by this operator, through the inclusion of the gradient and the Laplacian, leads to performance improvement, as shown in Table \ref{tab:res-gradlapl}. These additional representations of features that already satisfy the mass conservation constraint provide meaningful information to the decoder.

\begin{table}[h]
\centering
\begin{tabular}{lccc|c}
\multicolumn{5}{l}{{\footnotesize \textit{\textsc{VESSEL}}} dataset} \\[1pt] 
\Xhline{2pt}
{\footnotesize \textbf{Model}} & {\footnotesize \textbf{Loss}} &{\footnotesize \textbf{VM-SMAE}} & {\small\textbf{P-SMAE} }& {\footnotesize \textbf{Total-SMAE} }\\
\hline
\hline
\multicolumn{5}{l}{{\small \underline{\emph{Reference}}}} \\[1pt] 
\hline
\rowcolor{cyan!15}
{\small \textit{\textsc{GraphTransformer}}} & {\small$\mathcal{L}_{\text{sup}} + \mathcal{L}_{\text{PDE}} + \mathcal{L}_{\text{MASS,channels}}$ }& 0.195 & 0.317  &  \textbf{0.512}\\
\rowcolor{green!15}
{\small \textit{\textsc{GraphMamba}}}     & {\small $\mathcal{L}_{\text{sup}} + \mathcal{L}_{\text{PDE}} + \mathcal{L}_{\text{MASS,channels}}$ }& 0.223 & 0.355 & 0.578  \\
\hline
\multicolumn{5}{l}{{\small \underline{\emph{Without Grad-Lapl Graph Convolution}}}} \\[1pt] 
\hline
\rowcolor{cyan!15}
{\small \textit{\textsc{GraphTransformer}}} & {\small $\mathcal{L}_{\text{sup}} + \mathcal{L}_{\text{PDE}}+ \mathcal{L}_{\text{MASS,channels}}$ } & 0.195 & 0.330 & 0.525 \\
\rowcolor{green!15}
{\small \textit{\textsc{GraphMamba}}}     & {\small $\mathcal{L}_{\text{sup}} + \mathcal{L}_{\text{PDE}}+ \mathcal{L}_{\text{MASS,channels}}$ } &0.226 & 0.364 & 0.590 \\
\hline
\end{tabular}
\caption{Effect of the \emph{Grad-Lapl Graph Convolution} on the {\footnotesize \textit{\textsc{VESSEL}} dataset.  VM stands for velocity magnitude and P for pressure contributions.}}
\label{tab:res-gradlapl}
\end{table}

\subsection{Multi-fidelity evaluation}
We have explicitly built a multi-fidelity model by decomposing the learning task into two stages, where the first network is constrained to predict the Stokes solution. This design choice requires introducing additional parameters for the $\text{NN}_{ST}$. We therefore evaluate the effectiveness of the proposed multi-fidelity pipeline by comparing it with an alternative approach in which these additional parameters are instead employed to construct a single, larger network that directly learns the Navier–Stokes solution starting from the 1D Stokes approximation. With respect to reference networks with the hyperparameters described in Section \ref{sec:hyperparam}, we increase the hidden dimension to $105$ for {\small \textit{\textsc{GraphMamba}}} and to $126$ for {\small \textit{\textsc{GraphTransformer}}}.\\
As reported in Table \ref{tab:res-single}, despite relying on a smaller latent representation, the multi-fidelity framework shows better results. In particular, the multi-fidelity {\small \textit{\textsc{GraphTransformer}}} and {\small \textit{\textsc{GraphMamba}}} improve the Total-SMAE by $8.5\%$ and $3.5\%$, respectively, compared to their single-fidelity version. Leveraging the relationship between the Stokes and Navier–Stokes solutions through two coupled networks is an effective approach and represents a further direction for incorporating physical priors into the model design.\\
Moreover, low-fidelity Stokes data are cheaper to obtain. Indeed, in our solver, the Stokes solution is first computed as an initial guess for the iterative Navier–Stokes non-linear solver. Without any additional computational cost, we already have extra supervised data available for training.

\begin{table}[h]
\centering
\begin{tabular}{lccc|c}
\multicolumn{5}{l}{{\footnotesize \textit{\textsc{VESSEL}}} dataset} \\[1pt] 
\Xhline{2pt}
{\footnotesize \textbf{Model}} & {\footnotesize \textbf{Loss}} &{\footnotesize \textbf{VM-SMAE}} & {\small\textbf{P-SMAE} }& {\footnotesize \textbf{Total-SMAE} }\\
\hline
\hline
\multicolumn{5}{l}{{\small \underline{\emph{Reference}}}} \\[1pt] 
\hline
\rowcolor{cyan!15}
{\small \textit{\textsc{GraphTransformer}}} & {\small$\mathcal{L}_{\text{sup}} + \mathcal{L}_{\text{PDE}} + \mathcal{L}_{\text{MASS,channels}}$ }& 0.195 & 0.317  &  \textbf{0.512}\\
\rowcolor{green!15}
{\small \textit{\textsc{GraphMamba}}}     & {\small $\mathcal{L}_{\text{sup}} + \mathcal{L}_{\text{PDE}} + \mathcal{L}_{\text{MASS,channels}}$ }& 0.223 & 0.355 & 0.578  \\
\hline
\multicolumn{5}{l}{{\small \underline{\emph{Single-fidelity model with only} $\text{NN}_{NS}$}}} \\[1pt] 
\hline
\rowcolor{cyan!15}
{\small \textit{\textsc{GraphTransformer}}} & {\small $\mathcal{L}_{\text{sup}} + \mathcal{L}_{\text{PDE}}+ \mathcal{L}_{\text{MASS,channels}}$ } & 0.214 & 0.346 & 0.560 \\
\rowcolor{green!15}
{\small \textit{\textsc{GraphMamba}}}     & {\small $\mathcal{L}_{\text{sup}} + \mathcal{L}_{\text{PDE}}+ \mathcal{L}_{\text{MASS,channels}}$ } &0.232 & 0.367 & 0.599 \\
\hline
\end{tabular}
\caption{Single- and multi-fidelity comparison on the {\footnotesize \textit{\textsc{VESSEL}} dataset. VM stands for velocity magnitude and P for pressure contributions.}}
\label{tab:res-single}
\end{table}

\subsection{Computational cost analysis}
The {\small \textit{\textsc{GraphTransformer}}} models achieve the best overall performance. The attention mechanism is particularly effective at capturing and combining both local and global relationships in the considered test cases. This comes at the cost of a significantly higher memory consumption. We have also proposed {\small \textit{\textsc{GraphMamba}}} as a more efficient alternative, which is still able to learn non-local dependencies.\\
To perform a fairer comparison between the two architectures, and to reduce biases related to memory usage, we also consider two larger Mamba configurations, namely {\small \textit{\textsc{Medium}}} and {\small \textit{\textsc{Large}}}. Compared to the reference hyperparameters (Section~\ref{sec:hyperparam}), the {\small \textit{\textsc{Medium}}} model has $(2,3)$ processing steps, hidden dimensions of $(60,105)$, and state dimensions of $(60,90)$; while the {\small \textit{\textsc{Large}}} model employs $(2,2)$ processing steps, hidden dimensions of $(75,150)$, and state dimensions of $(60,120)$.\\

\begin{figure}[ht] 
    \centering
    \includegraphics[width=0.8\textwidth]{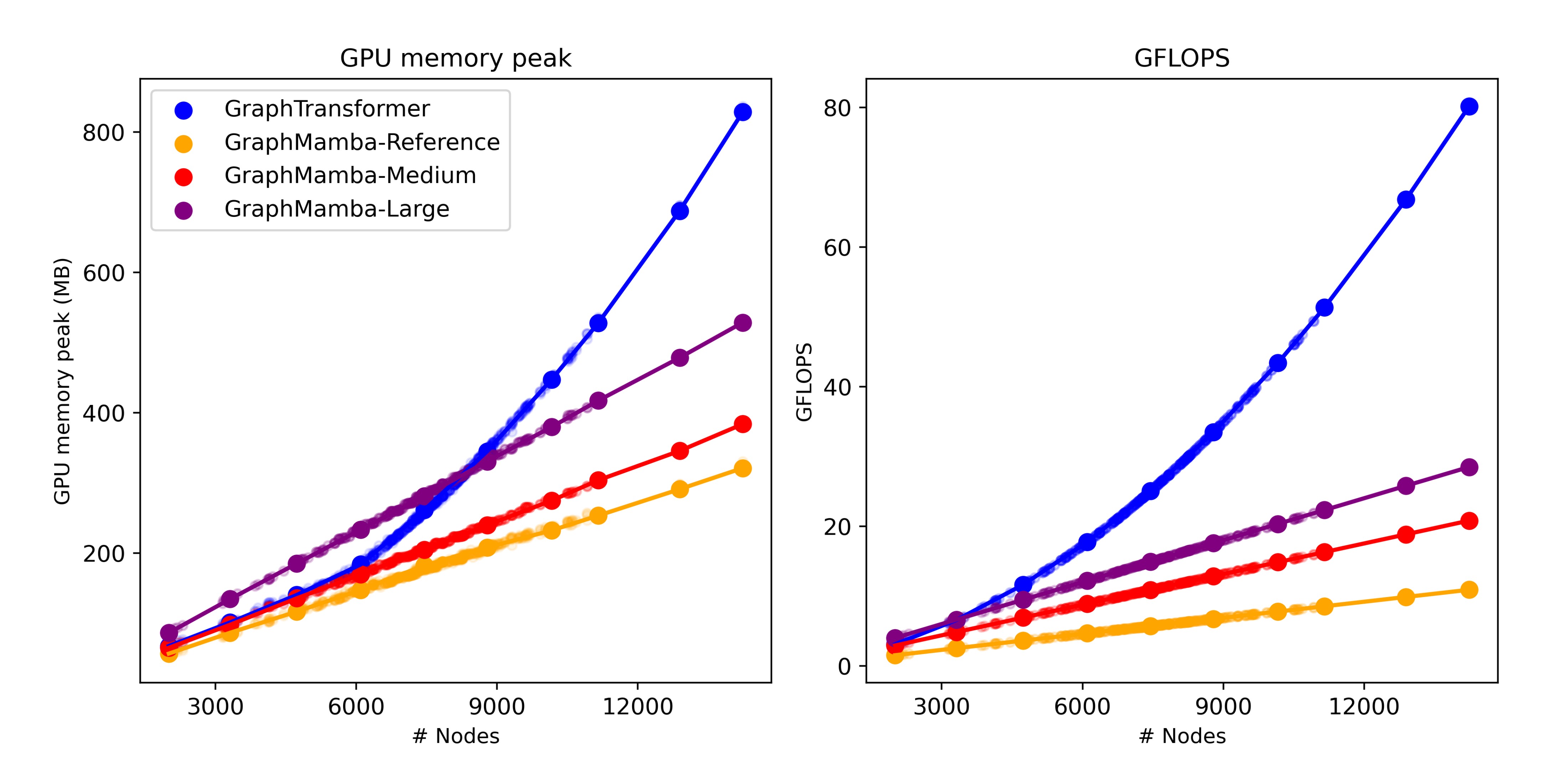}
    \caption{GPU memory peaks and the GFLOPs with respect to the number of nodes when evaluating the test set of the {\footnotesize \textit{\textsc{VESSEL}}} dataset for four different models.}
    \label{fig:gpu-mamba}
\end{figure}

In Figure \ref{fig:gpu-mamba}, we report GPU memory peaks and GFLOPs with respect to the number of nodes. These results are computed when evaluating the test set of the {\footnotesize \textit{\textsc{VESSEL}}} dataset (700 samples). As expected, we observe a quadratic behavior for the Transformer and a linear one for {\small \textit{\textsc{GraphMamba}}}. Despite improving the number of learnable parameters, as reported in Table \ref{tab:bigger-mamba}, {\small \textit{\textsc{GraphTransformer}}} still represents the better option to learn Navier-Stokes solutions in geometries represented with $\sim\!7500$ nodes from an accuracy point of view.

\begin{table}[H]
\centering
\begin{tabular}{lccc|c|c}
\multicolumn{6}{l}{{\footnotesize \textit{\textsc{VESSEL}}} dataset} \\[1pt] 
\Xhline{2pt}
{\footnotesize \textbf{Model}} & {\footnotesize \textbf{Loss}} & {\footnotesize \textbf{VM-SMAE}} & {\small\textbf{P-SMAE} }& {\footnotesize \textbf{Total-SMAE} } & Params (k)\\
\hline
\hline
\multicolumn{5}{l}{{\small \underline{\emph{Reference}}}} \\[1pt] 
\hline
\rowcolor{cyan!15}
{\small \textit{\textsc{GraphTransformer}}} & {\small$\mathcal{L}_{\text{sup}} + \mathcal{L}_{\text{PDE}} + \mathcal{L}_{\text{MASS,channels}}$ }& 0.195 & 0.317  &  \textbf{0.512} & 592\\
\rowcolor{green!15}
{\small \textit{\textsc{GraphMamba}}}     & {\small $\mathcal{L}_{\text{sup}} + \mathcal{L}_{\text{PDE}} + \mathcal{L}_{\text{MASS,channels}}$ }& 0.223 & 0.355 & 0.578 & 669 \\
\hline
\multicolumn{6}{l}{{\small \underline{\emph{Bigger Mamba SSM to match the Transformer computational cost }}}} \\[1pt] 
\hline
\rowcolor{green!15}
{\small \textit{\textsc{GraphMamba - Medium}}}     & {\small $\mathcal{L}_{\text{sup}} + \mathcal{L}_{\text{PDE}}+ \mathcal{L}_{\text{MASS,channels}}$ } & 0.216 & 0.357 & 0.573 & 1298 \\
\rowcolor{green!15}
{\small \textit{\textsc{GraphMamba - Large}}}     & {\small $\mathcal{L}_{\text{sup}} + \mathcal{L}_{\text{PDE}}+ \mathcal{L}_{\text{MASS,channels}}$ } & 0.205   & 0.336 & 0.541 & 1765\\
\hline
\end{tabular}
\caption{Error metrics and total parameters for the reference models and bigger {\small \textit{\textsc{GraphMamba}}} architectures. VM stands for velocity magnitude and P for pressure contributions.}
\label{tab:bigger-mamba}
\end{table}

\section{Discussion and Conclusion}
We have explored a multi-fidelity pipeline for learning steady Navier-Stokes solutions in non-parametrized 2D geometries, which exhibit a high level of geometric variability, especially in the {\footnotesize \textit{\textsc{VESSEL}}} dataset. The learning process is guided by passing through low-fidelity Stokes approximations before reaching the final high-fidelity solution. This represents a possibility to rely on low-fidelity data that are obtained without heavy additional computation. The advantages of relying on both low- and high-fidelity data are consistent with what is reported in \cite{howard2023multifidelity} on different problems with different physics. This strategy can be naturally extended to unsteady problems by using the prediction at the previous time step as an initial approximation for the next one. More generally, one can imagine a complete pipeline in which the model first learns the steady Navier–Stokes solution starting from Stokes, then considers this information to initialize the first time step of the unsteady problem, following directions from \cite{suk2025deep}, and finally advances the solution from one time step to the next.\\

Through the proposed \emph{encoding - processing - physics informed decoding} pipeline, physical constraints are introduced inside the architecture itself, guiding the model toward regular and physically consistent solutions. Combining mathematical knowledge within the deep learning architectures has been shown to improve performance. Thanks to the freedom provided by the numerical derivative operators (Section~\ref{wlsq}), we can easily introduce physical biases into the model, not only by enforcing the PDE residual on the final output, but also by enriching the latent representation itself. This is achieved by simple matrix multiplication with a node-defined field. With the \emph{Grad-Lapl Graph Convolution}, we provide additional physically meaningful information by computing the gradient and the Laplacian of selected hidden features. Currently, this is done only in the \emph{Decode} stage, but future work could explore the effect of applying such physics-informed graph convolutions also in earlier stages of the pipeline.\\
Moreover, this approach could naturally be extended within the multi-fidelity framework. For instance, in the Stokes equation, the pressure balances the diffusive term that arises from the velocity Laplacian, while in the Navier--Stokes equations, there is also the non-linear convective term. Concatenating the Laplacian and the convective term computed from the Stokes output into the $\text{NN}_{NS}$ input could therefore be informative when predicting $p_{NS}$. Even if sufficiently expressive architectures may, in principle, learn such operators implicitly, this strategy could explicitly guide the learning process using physical information that is well known and structured.\\

The Mamba architecture represents a valid alternative to Transformers for capturing non-local patterns in these domains. Although it was originally designed for sequential data, we can apply it to graph-structured data by relying on a novel unsupervised procedure to order the nodes. With Mamba, the results show that we reduce the computational costs, although the accuracy is lower than with Transformers. Despite this, predictions remain relevant and satisfactory. The added value of the Mamba architecture is expected to become more evident on larger domains, provided that a sufficiently large state is employed to capture and store global information.\\

We have identified some limitations of our approach that can be addressed in future work. Concerning the loss function, boundary conditions could be imposed in a weak form rather than being hardly enforced. In addition, the current loss function includes multiple terms, and the choice of their weights can be further improved. In this respect, approaches based on the conjugate kernel \cite{howard2024conjugate} could provide a way to dynamically adapt the relative importance of each term during training. It could be interesting to investigate whether first focusing the training on the Stokes network, to obtain a cleaner low-fidelity approximation, can lead to improved global performance.\\
By looking at the error maps over the different geometries, we have identified regions that present larger errors. Geometries with high curvature or strong restrictions show, on average, higher errors than domains with straighter branches. The velocity is well predicted along the centerline, whereas larger errors appear just off the centerline. This is reasonable, as we provide more reliable information right on the centerline, thanks to the 1D Stokes approximation. Moreover, when there is an immediate restriction or a new branch, the velocity exhibits high gradients with small \emph{jets} that are difficult to capture. Pressure prediction can be improved in \emph{impact regions}, either at bifurcation points or in high-curvature turns.\\
These observations are consistent from a fluid dynamics point of view, as these are regions where the solution presents high gradients and complex patterns.
We can further guide the network relying on this physical insight to better encode these regions. For instance, we could impose larger attention at bifurcation or \emph{impact regions}; better encode the geometrical features of the geometry starting from the centerline curvature, or encode the entire point cloud with an autoencoder-like shape model. One could also introduce additional states in the Mamba architecture that focus only on specific regions of the domain and then combine them with the global state to exchange the captured information.\\
All of these directions directly support the main motivation of this work, which is to incorporate physical principles into the learning process to better constrain it and allow better generalization capabilities. This framework can be readily applied to other physical applications modeled by PDEs, and it represents a first step that will be extended to more complex three-dimensional domains in future work.

\section*{Code and data availability}
The code to generate the data and to perform the training of the presented models will be made available on Codeberg upon acceptance. 
\section*{CRediT authorship contribution statement}
\textbf{Francesco Songia:} Writing – original draft, Visualization, Validation, Software, Methodology, Investigation, Formal analysis, Data curation, Conceptualization. \textbf{Raoul Sallé de Chou:} Writing – review \& editing, Supervision, Methodology, Software, Conceptualization. \textbf{Hugues Talbot:} Writing – review \& editing, Supervision, Methodology, Conceptualization.  \textbf{Irene E. Vignon-Clementel:} Writing – review \& editing, Supervision, Methodology, Project administration, Funding acquisition, Conceptualization.
\section*{Declaration of competing interest}
The authors declare that they have no known competing financial interests or personal relationships that could have appeared to influence the work reported in this paper.
\section*{Acknowledgments}
We acknowledge funding from the European Research Council (ERC) under the European Union’s Horizon 2020 research and innovation program (Grant agreement No. 864313)
\section*{Declaration of generative AI and AI-assisted technologies in the manuscript preparation process}
During the preparation of this work, the authors used ChatGPT-OpenAI in order to rephrase some paragraphs. After using this tool, the authors reviewed and edited the content as needed and take full responsibility for the content of the published article.

\bibliographystyle{elsarticle-num} 
\bibliography{refs.bib}

\clearpage

\appendix
\section{Data}
\label{app1-data}
\subsubsection{Synthetic datasets}
{\footnotesize \textit{\textsc{VESSEL}}}. To create a diverse set of vascular-like geometries, a set of base shapes is first drawn manually. We can think about configurations resembling the letters 'X', 'Y', 'H', and 'J'. New shapes are then generated by applying controlled deformations to these base geometries. 
Three main types of deformations are considered: (1) \textit{random perturbations}, where selected boundary control points are displaced by random vectors $\Delta \mathbf{x}$ within a prescribed magnitude, introducing stochastic irregularities; (2) \textit{elastic deformations}, obtained through a smooth radial basis function interpolation that allows coherent bending or stretching. On a set of automatically identified control points $\{\mathbf{c}_i\}_{i=1}^{k}$, prescribed displacements are assigned, and the deformation of any internal point $\mathbf{x}$ is given by
\[
\boldsymbol{\phi}(\mathbf{x}) = \mathbf{x} + \sum_{i=1}^{k} w_i \exp\! \ (  -\|\mathbf{x}-\mathbf{c}_i\|^2 / \sigma^2);
\]
and (3) \textit{mirror transformations}, where the resulting shapes are optionally mirrored along one or both axes to further increase geometric diversity. This procedure yields a wide range of synthetic vascular geometries with controlled deformation magnitude and type.\\
The variability among the different shapes allows the learning process to generalize across domains. Moreover, there are also very simple shapes, such as the horizontal ones from the 'J' group, which exhibit simpler flow and pressure patterns. These shapes are easily learned and display features that recur in other regions of more complex geometries.\\

{\footnotesize \textit{\textsc{CYLINDER}}}. To generate diverse domains within this dataset, the tube dimensions, as well as the position and size of the cylinder obstacle, are randomly varied.\\

\subsubsection{Reference CFD solutions and initial approximation}
\label{ref-cfd}
Starting from binary images of each geometry, \texttt{gmsh} \cite{geuzaine2009gmsh} is used to generate a finite element mesh for the domain. The Python library \texttt{FEniCS} \cite{alnaes2015fenics}, which is based on the finite element method, is used to solve the stationary Stokes and the stationary Navier–Stokes equations in each domain. These fields are the reference solution, and they are used in training for the supervised term and in the evaluation. \\

We also vary the boundary conditions. For both datasets, the left boundaries are generally treated as inlets and the right ones as outlets. Within the {\footnotesize \textit{\textsc{VESSEL}}} dataset, the different geometries allow for distinct inlet-outlet configurations: the 'X'- and 'H'-like domains have two inlets and two outlets, the 'Y' shapes have one inlet and two outlets, and the 'J' shapes have only a single inlet and outlet. For each geometry, we consider all combinations of balanced or unbalanced inlet flows and equal or different outlet pressures. Unbalanced inlet flows are generated by redistributing a fixed total flow according to a randomly chosen ratio $\gamma \in [0.25, 0.75]$, while different outlet pressures are imposed by setting one outlet to zero and assigning the other a value randomly sampled within $[15, 30]~\mathrm{Pa}$. Not all combinations are feasible for every shape: 'X' and 'H' domains can accommodate all possible combinations of inlet flow and outlet pressure, 'Y' shapes only support configurations with balanced inlet flow and equal or different outlet pressures. 'J' shapes are limited to a single inlet flow and outlet pressure configuration. In the final dataset, these boundary conditions are varied to ensure a wide range of scenarios.\\
For the {\footnotesize \textit{\textsc{CYLINDER}}} dataset, the outlet pressure is always set to zero, while the inlet boundary condition is defined through a parabolic velocity profile whose maximum value can vary across simulations.\\

Centerlines are automatically identified from the point cloud using an algorithm based on the signed distance function from the wall boundaries. The 1D Stokes equations are then solved along these centerlines for flow and pressure, with the domain represented as a network of nodes and edges: pressures are assigned to the nodes, while flows are associated with the edges. The connectivity of the network is encoded in a matrix $\mathbf{A}$, where each row corresponds to an edge and columns to nodes; entries $+1$ and $-1$ indicate the nodes connected by the edge and the flow direction. Poiseuille resistance is used to relate pressure drops to flows along each edge, with $R = 12 \mu L / S^3$ for an edge of length $L$ and cross-sectional area (distance) $S$. Node pressures $\mathbf{p}$ are obtained by solving the linear system $\mathbf{A}^\top \mathbf{C} \mathbf{A} \, \mathbf{p} = \mathbf{b}$, where $\mathbf{C}$ is a diagonal matrix of inverse resistances and $\mathbf{b}$ enforces inlet and outlet conditions. Finally, flows along the edges are computed from node pressures via $\mathbf{q} = \mathbf{C} \, \mathbf{A} \, \mathbf{p}$.\\

To use the 1D Stokes results as input features for the neural networks, it is necessary to extend the solution from the centerline to the nodes of the entire 2D mesh. This is achieved by introducing \emph{sections}, which are lines (planes in 3D) orthogonal to the centerline at each edge. Along each section, the velocity profile is assumed to be parabolic, with zero velocity at the vessel walls and a maximum velocity $v_{\mathrm{max}}$ at the center. From the 1D flow, we compute $v_{\mathrm{max}}$. Pressure is assumed constant within each section, equal to the pressure computed at the corresponding node in the 1D model. Once the velocity and pressure are defined on all sections, an iterative interpolation procedure is applied to propagate these values between consecutive sections, thereby covering the entire 2D domain. In this way, every point in the mesh is assigned a physically consistent velocity and pressure, providing a complete 2D field derived from the 1D approximation.\\

\section{Additional visualization}
\label{app2-addit-visualization}
\begin{figure}[H] 
    \centering
    \includegraphics[width=0.95\textwidth]{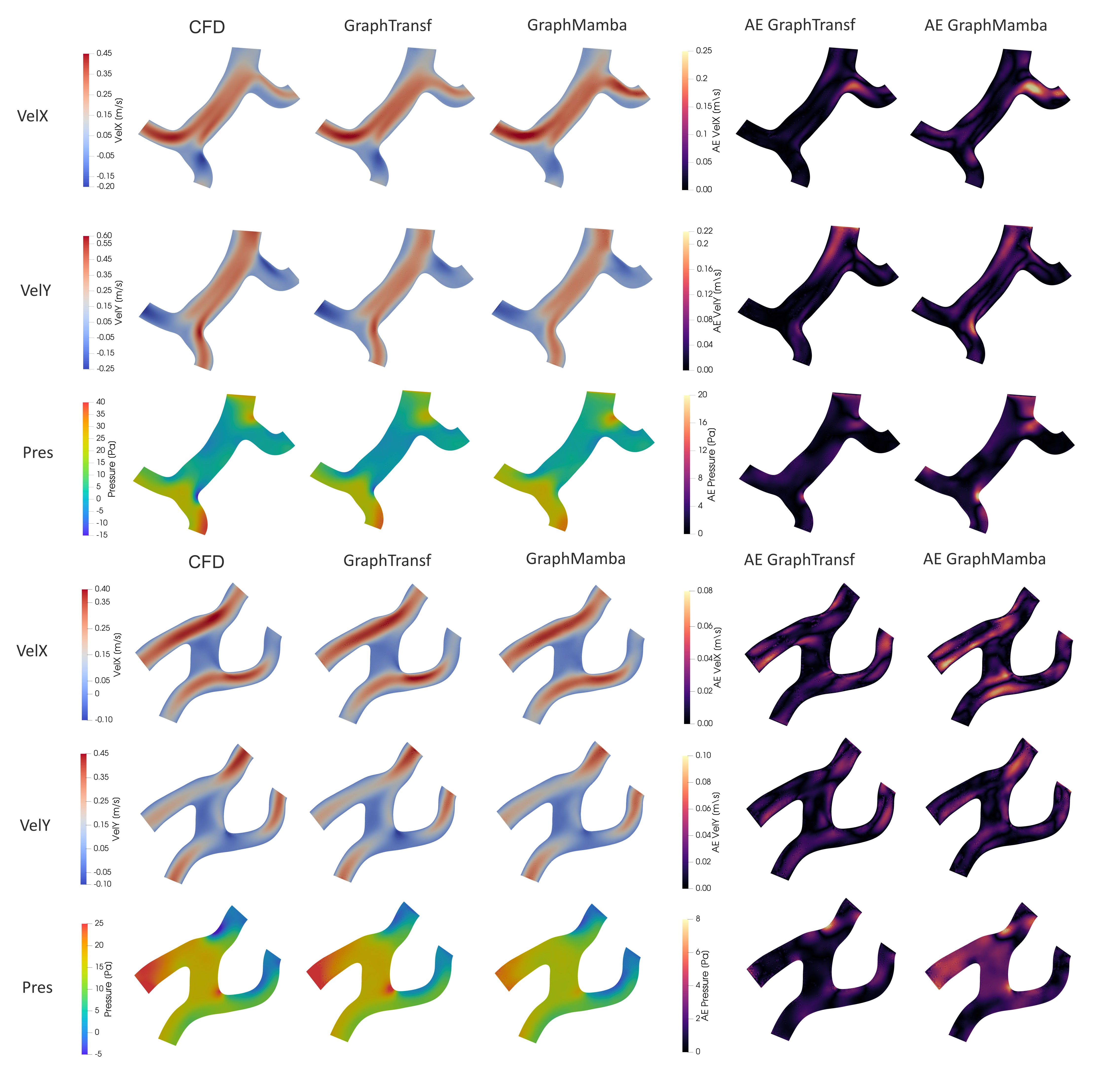}
    \caption{CFD ground truth and predictions of {\small \textit{\textsc{GraphTransformer}}} (2nd column) and of {\small \textit{\textsc{GraphMamba}}} (3rd column) on the {\footnotesize \textit{\textsc{VESSEL}}} test dataset. They both refer to the best loss configuration with mass conservation enforced on the \emph{channels}. On the right, there are corresponding absolute error maps.}
    \label{fig-app2-ves}
\end{figure}

\begin{figure}[H] 
    \centering
    \includegraphics[width=0.95\textwidth]{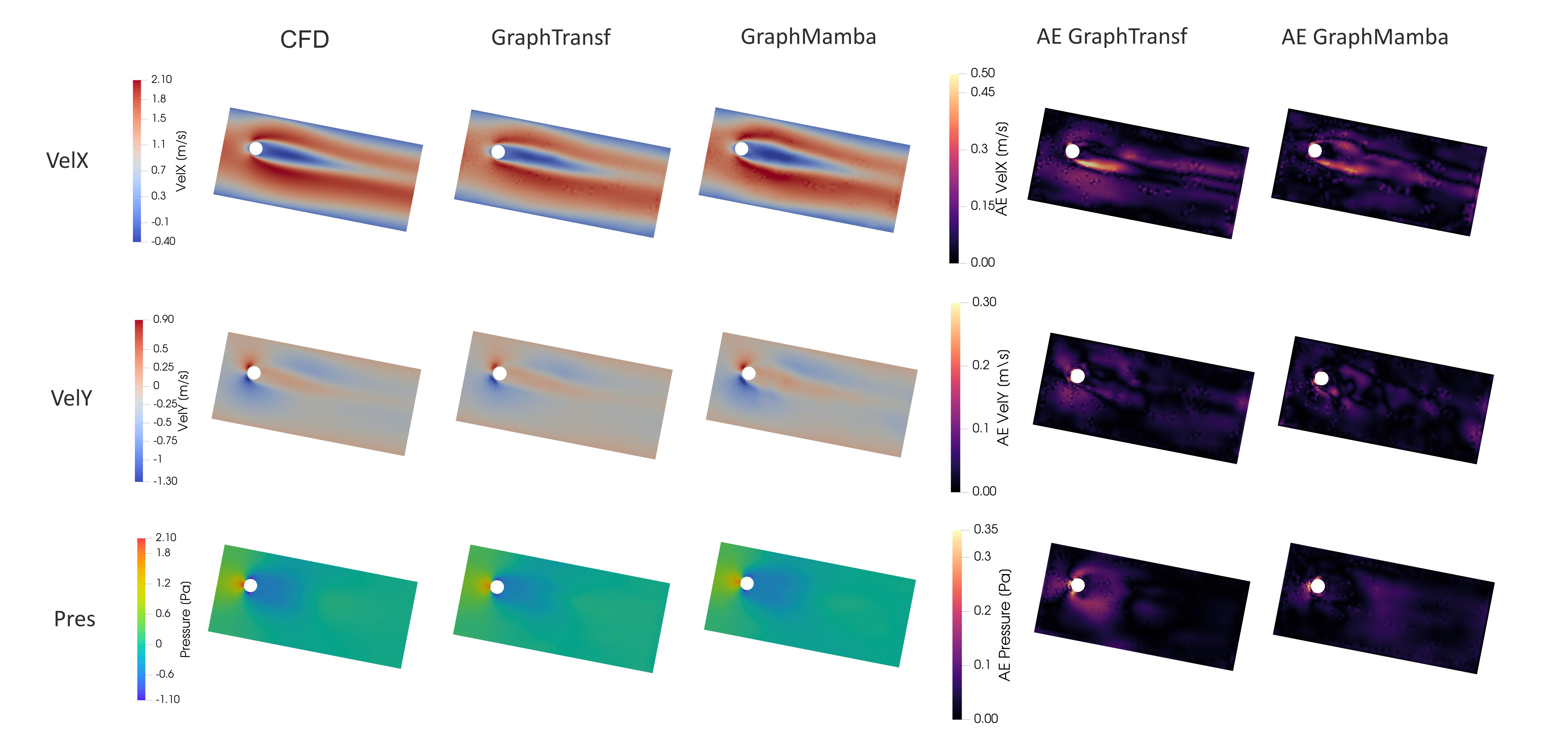}
    \caption{CFD ground truth and predictions of {\small \textit{\textsc{GraphTransformer}}} (2nd column) and of {\small \textit{\textsc{GraphMamba}}} (3rd column) on the {\footnotesize \textit{\textsc{CYLINDER}}} test dataset. They both refer to the best loss configuration with mass conservation enforced on the \emph{channels}. On the right, there are corresponding absolute error maps.}
    \label{fig-app2-cyl}
\end{figure}

\end{document}